\documentclass[a4paper,11pt]{article}

\usepackage[margin=1in]{geometry} 

\newenvironment{acknowledgments}{%
  \section*{Acknowledgments}%
}{}

\usepackage[utf8]{inputenc}  
\usepackage{array}
\usepackage{makecell}   
\usepackage{graphicx}   
\usepackage{dcolumn}
\usepackage{bm}
\usepackage{graphicx}
\usepackage{booktabs}
\usepackage{xcolor}
\usepackage{xfrac}
\usepackage{mathtools}
\usepackage{amsmath}
\usepackage{amssymb}
\usepackage{comment}
\usepackage{siunitx}
\usepackage{nameref} 
\usepackage[colorlinks=true,linkcolor=blue,citecolor=blue,urlcolor=blue]{hyperref}
\usepackage{soul}
\usepackage{subfigure}
\usepackage{tikz}
\usepackage{tabularx}
\usepackage{multirow}
 
\renewcommand{\selectlanguage}[1]{}

\usepackage[mathlines]{lineno}

\usepackage{amsmath}
\usepackage{amsthm}
\usepackage{amssymb}
\usepackage{lineno}
\usepackage{subcaption}  

\usepackage{dutchcal}

\usepackage{placeins}

\usepackage{etoolbox}

\usepackage{graphicx}
\usepackage{bm}
\usepackage{lineno}
\usepackage{caption}

\usepackage{booktabs}

\usepackage{multirow}

\usepackage{comment}
\usepackage{xargs}                      

\hypersetup{
	unicode,
	colorlinks=true,
	linkcolor=blue,
	filecolor=magenta,      
	citecolor=blue,
	urlcolor=cyan,
	pdfauthor={ M. A. Mendez},
	pdftitle={Linear and Nonlinear Dimensionality Reduction from Fluid Mechanics to Machine Learning},
	pdfsubject={Linear and Nonlinear Dimensionality Reduction from Fluid Mechanics to Machine Learning},
	pdfkeywords={article, template, simple},
	pdfproducer={LaTeX},
	pdfcreator={pdflatex}
	colorlinks=true,
	citecolor=blue,
	linkcolor=blue,
	filecolor=magenta,      
	urlcolor=blue,
}

\usepackage[most]{tcolorbox}
\usepackage{booktabs}
\usepackage{amsmath}
\usepackage{wrapfig}

\usepackage{mathtools}
\tcbset{enhanced,colback=cyan!2!white,colframe=cyan!90!white,coltitle=blue!20!black,fonttitle=\bfseries}

\usepackage{listings}
\usepackage{hyperref}
\hypersetup{
	colorlinks=true,
	linkcolor=blue,
	filecolor=magenta,      
	citecolor=blue,
	urlcolor=cyan,
}

\usepackage[sort&compress,authoryear,round]{natbib}
\usepackage{algorithm, algpseudocode} 
\usepackage{mathrsfs} 

\usepackage{siunitx}
\newcolumntype{Z}[1]{S[scientific-notation=fixed, fixed-exponent=#1, table-format=1.2]}

\title{\LARGE \textbf{Regime Maps for Sloshing in Horizontal Cylindrical \\Tanks Under Vertical Acceleration}}

\author{
  Francisco Monteiro$^{1,2}$\thanks{Corresponding author: francisco.monteiro@vki.ac.be} \and
  Tommaso De Maria$^{1, 3}$ \and
  Samuel Ahizi$^{1,2}$ \and
  Ramòn Abarca$^{4}$ \and
  Giuseppe C.A. Caridi$^{1}$ \and
  Miguel A. Mendez$^{1,2,5}$
}

\date{%
\begingroup
\small
\textit{%
$^1$ Environmental and Applied Fluid Dynamics, von Karman Institute for Fluid Dynamics, Belgium \\
$^2$ Aerospace Engineering Research Group, Universidad Carlos III de Madrid, Leganés, 28911, Madrid, Spain \\
$^3$ SAAS, Universit{\'e} Libre de Bruxelles, Belgium \\
$^4$ Airbus Operations S.L., Spain \\
$^5$ Aero-Thermo-Mechanics Laboratory, École Polytechnique de Bruxelles, Université Libre de Bruxelles, Belgium}
\endgroup
}

\begin{document}
\maketitle
	
\begin{abstract}

Vertical sloshing in partially filled fuel tanks can significantly impact vehicle stability and structural integrity, particularly under harmonic accelerations near twice the sloshing natural frequency. In this regime, parametric resonance may arise, with nonlinear free-surface dynamics driving large-amplitude waves, interface break-up, and severe sloshing-induced mixing. In this work, we identify and characterize the distinct sloshing regimes associated with the lowest-frequency parametric instability, specifically when the external forcing frequency approaches twice the lowest natural frequency. Experiments were conducted in a transparent cylindrical tank with diameter $D = \SI{134.5}{\milli\meter}$ and length $L = \SI{336.3}{\milli\meter}$. This work presents a data-driven approach for regime identification and classification that relies solely on high-speed video recordings and circumvents the need for interface tracking. The method combines prototype-based data labeling with dimensionality reduction via multiscale proper orthogonal decomposition (mPOD) and automatic kernel-based classification. The results are summarized in a dimensionless regime map across three fill ratios ($H_l/D \in [0.40; 0.67]$), where stable waves, longitudinal and transverse mode shapes, and mode-competition regimes are distinguished. The developed map provides a predictive tool for assessing sloshing-induced loads, supporting structural and operational optimization of fuel systems.
			
\vspace{7mm}
\noindent\textbf{Keywords:} Faraday waves, vertical sloshing regimes, parametric resonance, data-driven classification
\end{abstract}

\section{Introduction}

\begin{figure*}[t!]
\centering
\includegraphics[width=0.9\textwidth]{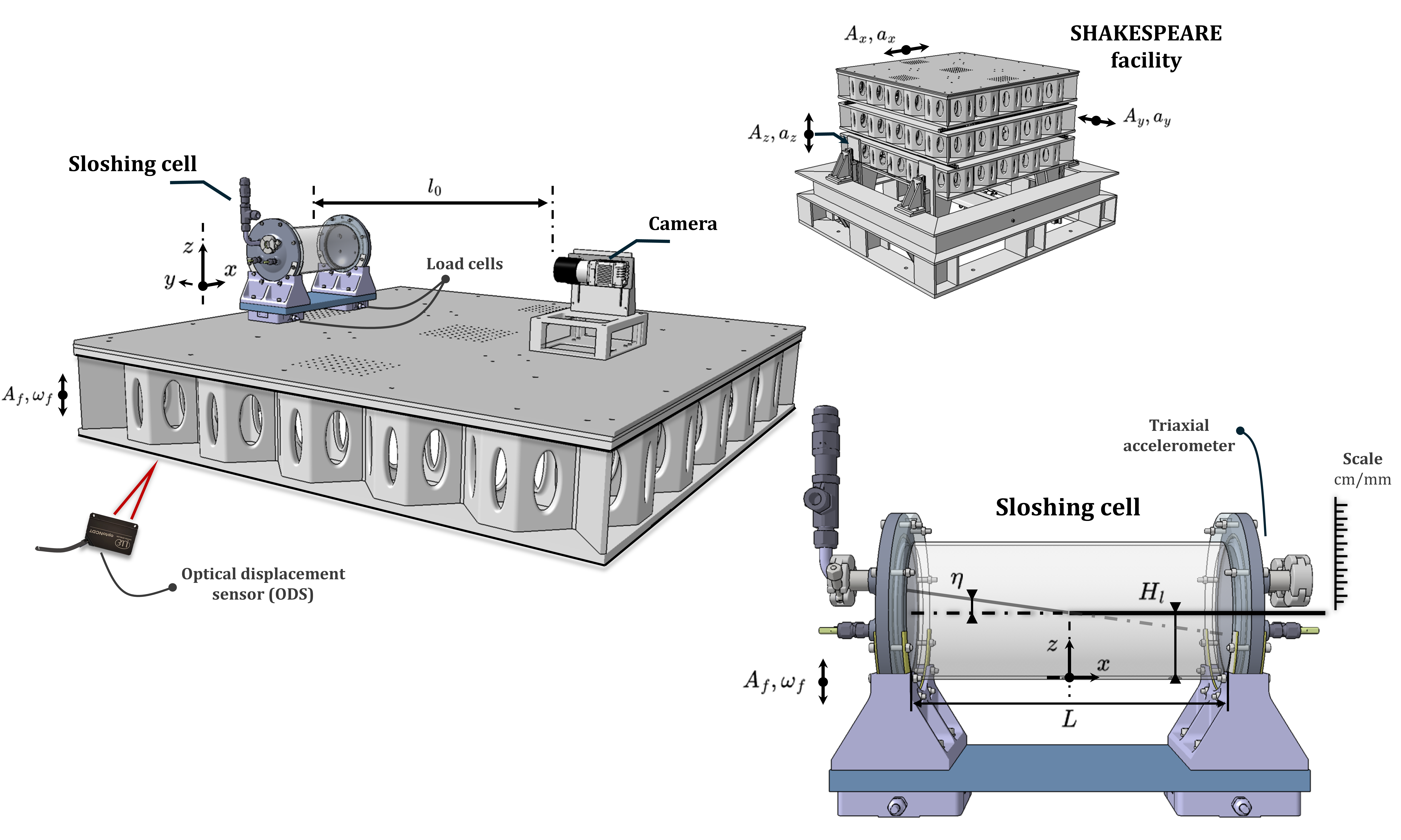}
\caption{Experimental setup integration in the SHAKESPEARE sloshing table facility. The sloshing cell comprises a PMMA horizontal cylindrical test section with an aspect ratio of $L/R = 5.0$, with aluminum cylindrical end domes. A high-speed camera acquires the free-surface displacement ($\eta$) under vertical accelerations ($a_z$), and the dynamic conditions are controlled through an optical displacement sensor and a triaxial accelerometer ($\boldsymbol{a}(t) = \left(a_x(t), a_y(t), a_z(t)\right)$).} 
\label{fig:experimental_setup}
\end{figure*} 
 
\begin{table*}[t!]
\caption{Experimental matrix summarizing the investigated conditions. Reported parameters include the liquid fill ratio $H_l/D$, the number of executed test points $N_\text{pts}$, the target sloshing mode natural frequency $\omega_{1,0}$ and damping ratio $\zeta_{1,0}$, and the explored excitation ranges: dimensionless vertical acceleration $A_f\omega_f^2/g$ (in units of $g$), dimensionless displacement amplitude $A_f/R$, and dimensionless forcing frequency $\omega_f/(2\,\omega_{1,0})$.}
\centering
\renewcommand{\arraystretch}{1.10}
\begin{tabular*}{\textwidth}{p{0.1\textwidth}@{\extracolsep{\fill}}|cccccc}
\toprule
$H_l/D$ & $N_\text{pts}$ & $\bar{\omega}_{1,0}$ [\unit{\radian \per \s}] & $\bar{\zeta}_{1,0}$ [-] & $\pm A_f\omega_f^2/g$ [-] & $A_f/R$ [-] & $\omega_f/(2\,\omega_{1,0})$ [-] \\
\midrule[1.25pt]
0.67 & 66 & $7.12 \pm 0.03$  & $0.015 \pm 0.002$ & $0.02 - 0.94$ & $0.02-0.58$ & $0.90-1.20$ \\
0.50 & 73 & $6.08 \pm 0.16$  & $0.004 \pm 0.001$ & $0.02 - 0.77$ & $0.02-0.62$ & $0.90-1.20$ \\
0.40 & 61 & $5.32 \pm 0.02$  & $0.006 \pm 0.002$ & $0.02 - 0.48$ & $0.02-0.52$ & $0.87-1.16$ \\
\bottomrule
\end{tabular*}
\label{tab:experimental_matrix}
\end{table*}

The dynamics of liquid sloshing have long been a concern in engineering systems that store large fluid volumes, such as chemical and power plants, aircraft and rocket propellant tanks, or cargo ships \citep{abramson_1981, kobayashi_study_1989}. Whether triggered by seismic events or vehicle accelerations, sloshing displaces the tank's center of mass \citep{PRF_walking_coffe}, representing a critical source of dynamic loading that can compromise stability, maneuverability, or structural integrity \citep{dodge_2000}. Accurately predicting and mitigating such effects requires a detailed analysis of the fluid behavior under various conditions \citep{faltinsen_sloshing_2009}, which is primarily dictated by tank geometry, excitation direction, and fill level \citep{ibrahim_2005,monteiro_microgravity, Kalinichenko_viscosity_2018}.

Research on sloshing dynamics has historically focused on upright cylindrical tanks subjected to lateral excitation -- a configuration of particular relevance to early space programs, where lateral sloshing was a central concern during the ascent phase \citep{arndt_2012, Vanforeest_2014, marques_assimilation_model, marques_experimental_2023}. A key outcome of this research was the development of a dimensionless forcing diagram that characterizes the dynamic regimes near the fundamental sloshing mode \citep{miles_1984}, where swirling and chaotic motions can emerge depending on the excitation-to-natural frequency ratio \citep{royon-lebeaud, hopfinger_baumbach_2009}.

No similar regime map exists for cylindrical horizontal tanks under vertical excitation — a configuration increasingly relevant in aviation, where cruise conditions expose fuel tanks to low-frequency, quasi-harmonic vertical accelerations from turbulence, gusts, or control inputs. While traditionally less studied, vertical sloshing has recently become a prominent research focus, notably within the SLOshing Wing Dynamics (SLOWD) initiative \citep{SLOWD_sloshing_2020, PRF_slowd_1, PRF_slowd_2, constantin_analysis_2021, constantin_nonlinear_2023} and the Hydrogen Aircraft Sloshing Tank Advancement (HASTA) project \citep{hasta2025}.

This renewed attention is partly driven by the risk of parametric resonance, which arises when the frequency of the vertical accelerations, $\omega_f$, approaches twice the natural frequency of a free-surface sloshing mode, $\omega_{m,n}$, leading to the onset of subharmonic Faraday waves \citep{faraday1831, rayleigh1916, ibrahim_recent_2015}. Here, the index $m$ is the longitudinal index (number of half-wavelengths along the cylinder axis) and $n$ is the transverse index (number of nodal diameters across the cross-section/radius).

The critical feature of vertical sloshing is that the excitation modulates the effective gravity, which normally provides the restoring force for free-surface oscillations. When this modulation interacts resonantly with the free-surface dynamics, parametric instability can occur. Though requiring much larger forcing amplitudes than in lateral sloshing \citep{dodge_1965, brand_parametrically_1965}, such resonance can be critical if not mitigated \citep{chang_nonlinear_2014, yu_baffles_experimental_2020}.

The onset of these waves can be described by the classic Mathieu's equation, with instability regions mapped in Ince–Strutt diagrams around subharmonic and harmonic frequencies \citep{benjamin_ursell_1954, miles_parametrically_nodate}. In practice, most of the concern is limited to the lowest frequency modes for which viscous damping is not particularly effective \cite{Woodward_1966}. Within the unstable regions, finite-amplitude and exponential growth waves can develop with mode competition arising due to overlapping instability bands \citep{dodge_1965, ciliberto_chaotic_1985}. Outside such regions, the liquid-free surface remains stable/flat \citep{el-dib_nonlinear_2001, frandsen_sloshing_2004}, except for a small response close to the driving frequency \citep{ciliberto_pattern_1984}. 

Literature on sloshing near parametric resonance has mostly focused on high aspect ratio rectangular tanks ($L \gg D$), where longitudinal modes dominate, and transverse motion is suppressed \citep{kalinichenko_breaking_2009, jiang_period_1998, hashimoto_violent_1988}. Research has largely examined the second $(m = 2, n=0)$ and third $(m = 3, n=0)$ longitudinal modes, including jet formation at wave crests, and categorized Faraday waves as regular, irregular, or breaking \citep{kayal_jet_2023, Kalinichenko_viscosity_2018}. Experiments in cylindrical and square geometries have revealed pure one-mode motion, mode competition with chaotic or periodic transitions, and rotational states among degenerate or interacting modes \citep{ciliberto_phenomenological_1985, Feng_Sethna_1989, GOLLUB_surface_1989, das_parametrically_2008, das_hopfinger_2009}. 

For horizontal cylindrical tanks, much of the prior work has focused on lateral and pitching excitations \citep{luo_cassini_experiments_2023, kobayashi_study_1989, grotle_2018, martin_lopez_estudio_2013} or on the numerical computation of natural frequencies \citep{hasheminejad_analytical_2017, han_semi-analytical_2021}. More recently, quasi-2D configurations with $L \ll D$ have been investigated \citep{colville_faraday_2025, saltari_experimental_2025}, to investigate the transverse dynamics while suppressing the longitudinal one (i.e., only modes with $m=0$ can be observed).

This work reports the first extensive experimental characterization of vertical sloshing of a horizontal cylindrical tank under gravity-dominated conditions near the parametric resonance $\omega_f \approx 2\,\omega_{1,0}$. A combination of modal decomposition, clustering, and support vector machine classification was used to identify nonlinear sloshing regimes from backlighting high-speed visualizations and organize these into a predictive map. 

The article is structured as follows. Section \ref{sec:investigated_conditions} presents the problem set with all the quantities of interest, and Section \ref{sec:Mathieu_eq} introduces the Mathieu equation and defines the onset conditions of the instability regions relevant to parametrically excited sloshing. Section \ref{sec2-exp_setup} overviews the experimental setup. Section \ref{sec3-regime_id} reports on the methodology, describing the image processing, clustering algorithm, and automatic classification tool used for identifying the sloshing regime. Section \ref{sec4-results} reports the experimental results with the natural frequency determination, viscous damping factor computation, and concludes with the neutral-stability boundary maps across multiple fill levels. Conclusions and perspectives are presented in Section \ref{sec5-conclusions}.


\section{Investigated Conditions} \label{sec:investigated_conditions}

The configuration of interest is a horizontal cylindrical tank of radius $R$ and cylindrical length $L$. The tank ends with spherical domes of radius $R_d$, whose centers are located at a distance $l_d$ from the flat cylinder--dome interface plane. The tank is filled to a height $H_l$ measured from the bottom with a liquid of density $\rho$ and dynamic viscosity $\mu$. 
The ullage volume is filled with air, whose properties do not play a role in the sloshing dynamics and are thus not considered further. The surface tension at the gas-liquid interface is denoted as $\sigma$.

The tank is subject to a harmonic forcing in the vertical direction, with a displacement law $z(t)=A_f \sin(\omega_f t)$, where $A_f$ and $\omega_f$ are the forcing amplitude and angular frequency, respectively. The vertical acceleration experienced by the tank then reads $a_z(t)=-g-A_f\omega^2_f\sin(\omega_f t)$, with $g = \SI{9.81}{\m \per \square \s}$ the gravitational acceleration. 


Denoting the interface dynamics by $\eta(\bm{x},t)$, where \mbox{$\bm{x}=(x,y) \in \mathbb{R}^2$} represents the in-plane (horizontal) coordinate vector, the problem depends on three geometrical parameters, one kinematic parameter, one operational parameter (the liquid fill level), and three dimensionless numbers governing the force balance:
\begin{equation}
\frac{\eta(\bm{x},t)}{R} = f\biggl(\frac{L}{R}, \frac{R_d}{R}, \frac{l_d}{R}, \frac{A_f}{R}, \frac{H_l}{D}, \text{Re, Fr, We}  \biggr)\,,
\end{equation} where \mbox{$\text{Re}=\rho [U] R/\mu$} is the Reynolds number, \mbox{$\text{We}=\rho [U]^2 R/\sigma$} is the Weber number and $\text{Fr}=[U]/\sqrt{g R}$ is the Froude number. Here, the reference velocity is taken as $[U] = (A_f\,\omega_f)$. Within the range of investigated conditions, surface tension and wetting effects can be considered negligible, since $\text{We}$ ranges from \( \mathcal{O}(10^1) \) to \( \mathcal{O}(10^3) \) and the ratio $\text{Bo}=\text{We}/\text{Fr}^2$, also known as Bond number \citep{jang_mechanical_BOND}, is of \(\sim \mathcal{O}(10^3) \). The Reynolds number lies between $\mathcal{O}(10^3)$ and $\mathcal{O}(10^4)$, and thus moderate turbulence is to be expected, while the Froude number remains below unity in all experiments, hence the investigated conditions are clearly in gravity-dominated conditions \citep{abramson_1981, dodge_2000}.

Sloshing regimes are traditionally analyzed in terms of the system’s natural frequency. For vertical sloshing, one usually considers the ratio $\omega_f / (2\,\omega_{1,0})$, which scales the forcing relative to the first longitudinal mode, since the first parametric resonance is expected to occur near $\omega_f \approx 2\,\omega_{1,0}$. However, analytical expressions for these natural frequencies exist only for flat-end cylindrical tanks (see \citet{hasheminejad_analytical_2017}), but not for the dome-ended geometries investigated here. Therefore, the natural frequency $\omega_{1,0}$ was experimentally identified from free-decay tests as described in Subsec. \ref{sec:methodology_nat_freq}. The corresponding results, presented in Subsec.~\ref{sec:nat_freq_experimental}, are significantly lower than those predicted by the flat-end cylinder formulation of \citet{hasheminejad_analytical_2017}, which are included in Appendix~\ref{app:natfreq_theory} for reference.

The experimental campaign was conducted at three fill ratios, $H_l/D = 0.40, 0.50$, and $0.67$. Table \ref{tab:experimental_matrix} summarizes the range of conditions $(A_f\omega_f^2/g,\, A_f/R,\,\omega_f/(2\,\omega_{1,0}))$ and the number of tests, $N_{pts}$, for each fill level. Within these ranges, sampling followed a two–stage approach. First, amplitude sweeps at a constant forcing frequency near the primary resonance, $\omega_f/(2\,\omega_{1,0}) \approx 1$, were performed to determine the critical amplitude at the transition between stable and unstable dynamics. Second, we carried out a randomized, space-filling exploration of the remaining domain to achieve near-uniform coverage.   


\section{Parametric Resonance and Instability Regions} \label{sec:Mathieu_eq}

Under vertical sloshing conditions, the imposed excitation modulates the effective gravity acting on the liquid mass. In the traditional inviscid, incompressible, and irrotational flow setting, linearizing the dynamic boundary condition of the interface allows to write the free surface elevation $\eta(\bm{x},t)$ as the superposition of sloshing eigenmodes: 
\begin{equation}
\eta(\bm{x},t)=\sum_{m,n} \eta_{m,n}(t) \xi_{m,n}(\bm{x})\,,
\label{modes}
\end{equation} where $\xi_{m,n}$ are the spatial mode shapes satisfying the linear eigenvalue problem for the tank geometry and $\eta_{m,n}(t)$ are the time-dependent modal amplitudes. Substituting this expansion into the governing equations and projecting onto each eigenmode yields, for every pair $(m,n)$, an uncoupled Mathieu-type oscillator of the form
\begin{align}
  \frac{d^{2}\eta_{m,n}(t)}{dt^{2}}
  & + 2\,\zeta_{m,n}\,\omega_{m,n}\,\frac{d\eta_{m,n}(t)}{dt} \nonumber \\
  & + \omega_{m,n}^{2}\Biggl(1 - \frac{A_f\,\omega_f^2}{g} \cos(\omega_f t)\Biggr)\,\eta_{m,n}(t)
  = 0, 
  \label{eq:mathieu_damped_d_dt}
\end{align} where $\omega_{m,n}$ and $\zeta_{m,n}$ denote the natural frequency and damping ratio of the $(m,n)$ mode. Although the tank motion is applied as a sinusoidal displacement \mbox{$z(t) = A_f \sin(\omega_f t)$}, the cosine form is adopted here for consistency with the canonical Mathieu equation \citep{CERDA_TIRAPEGUI_mathieu_eq, cerdA_mathieu_2}, noting that a phase shift of $\pi/2$ does not affect the predicted stability boundaries.

The instability regions of \eqref{eq:mathieu_damped_d_dt} were computed using Floquet stability theory, following the formulation summarized by ~\cite{ibrahim_2005}. The approach consists of recasting \eqref{eq:mathieu_damped_d_dt} as a first-order system with periodic coefficients and integrating over one forcing period $T = 2\pi/\omega_f$ to construct the monodromy matrix mapping the state of the system from any time $t$ to $t+T$. The Floquet multipliers --— that is, the eigenvalues of this matrix --— determine stability: the boundary of each Mathieu tongue corresponds to parameter combinations $(A_f/R, \,\omega_f/(2\,\omega_{m,n}))$ for which at least one multiplier lies on the unit circle, i.e. where the solution transitions from bounded to exponentially growing behavior.

The resulting neutral-stability boundaries define the so-called Mathieu tongues, which arise near integer subharmonic orders of the natural frequency, $\omega_f \approx 2 \omega_{m,n} / r$ with $r = 1, 2, 3, \ldots$. The principal instability region ($r = 1$) corresponds to the classical Faraday subharmonic resonance~\citep{faraday1831}.

It is worth noticing that the form in \eqref{eq:mathieu_damped_d_dt} also includes a term for viscous dissipation --- absent from the inviscid formulation of \citet{benjamin_ursell_1954} --- introduced through the linear damping coefficient $2\,\zeta_{m,n}\,\omega_{m,n}$, analogous to a classical second-order oscillator. This correction, proposed by \citet{brand_parametrically_1965}, was motivated by experiments showing that viscous effects shift the instability threshold. In the inviscid limit, instability regions extend from zero forcing amplitude ($A_f/R = 0$), whereas the inclusion of damping requires a finite critical amplitude for surface waves to develop ($A_f/R > 0$, at $\omega_f/(2\,\omega_{1,0}) = 1$).

In this work, we focus on the principal parametric-instability region associated with the $(m,n) = (1,0)$ mode. As this mode possesses the lowest natural frequency, it defines the first and widest accessible instability domain. Equation~\eqref{eq:mathieu_damped_d_dt} provides the theoretical basis for analyzing its modal dynamics, while the corresponding instability boundaries are presented in Section ~\ref{sec:sloshing_maps}.

\section{Experimental Setup} \label{sec2-exp_setup}

\begin{figure*}[t!]
\centering
\includegraphics[width=0.925\textwidth]{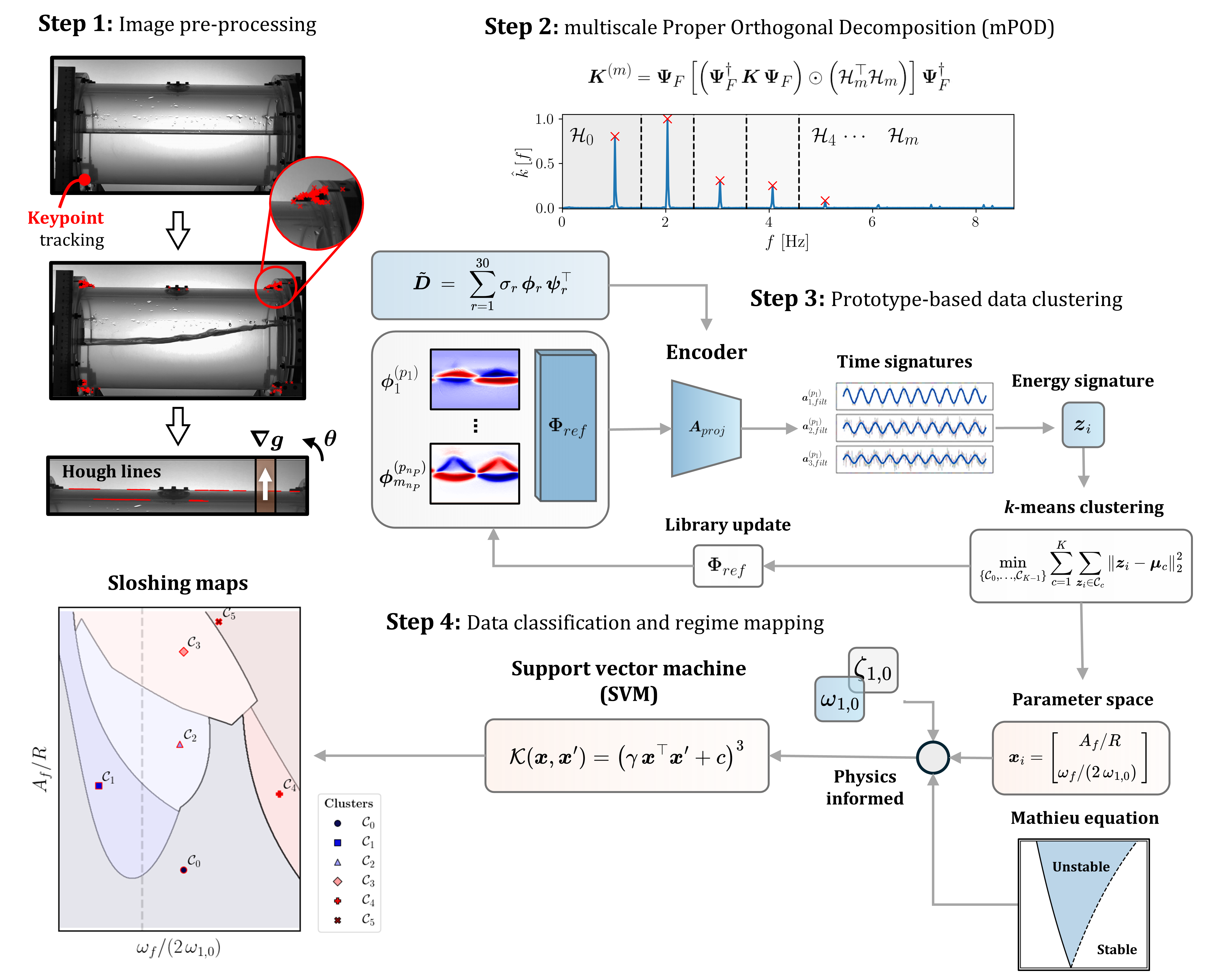}
\caption{\footnotesize \textbf{Flowchart of the proposed algorithm.} \textbf{Step 1:} acquired image sequences are pre-processed to correct for unwanted vibrations and tank tilt. \textbf{Step 2:} the mPOD is used to analyze the acquired snapshots between $t \in [0, 60]\,$ seconds. Data is then truncated to the first $n_R = 30$ mPOD modes. \textbf{Step 3:} a reference prototype library ($\boldsymbol{\Phi}_{ref}$) is created, and each truncated test case ($\tilde{\boldsymbol{D}}$) is projected into it to identify dominant patterns over time ($\boldsymbol{A}_{proj} \rightarrow \boldsymbol{z}_i$). Finally, unsupervised \( k \)-means clustering is used to label the data, and a non-linear multiclass SVM determines continuous decision boundaries between sloshing regimes (\textbf{Step 4}).} 
\label{fig:flowchart_methodology}
\end{figure*} 

The experimental setup, shown in Fig. \ref{fig:experimental_setup}, consists of a transparent polymethyl methacrylate (PMMA) reservoir with circular cross-section with diameter $D = \SI{134.5}{\milli\meter}$, length $L = \SI{336.3}{\milli\meter}$, and aspect ratio $L/R = 5.0$. Each end of the tank is closed by a spherical dome of radius $R_d = \SI{100.0}{\milli\meter}$ and depth \SI{25.0}{\milli\meter}, whose center of curvature lies $l_d = \SI{75.0}{\milli\meter}$ inside the cylindrical section of the tank.

The cell was partially filled with demineralized water, and air at ambient pressure occupied the ullage space. Tests were performed under near-isothermal conditions, with ambient temperature in the range $T_{amb} = [290,\,298]\,\si{\kelvin}$. Within the range of temperatures in the experiments, the average liquid properties, as computed from REFPROP \citep{refprop}, are ${\rho} = \SI{998}{\kilogram\per\cubic\meter}$, ${\mu} = \SI{981}{\micro\pascal\second}$, and ${\sigma} = \SI{72.6}{\milli\newton\per\meter}$.

Optical measurements using backlighting were performed with a \textit{JAI SP-12000MCXP4} high-speed camera at $\SI{70}{\Hz}$ equipped with a $f = \SI{75}{\milli\meter}$ objective lens. The focal ratio was fixed at \(f/4.0\). The camera was mounted on the sloshing table at a distance of \( l_0 = \SI{375}{\milli\meter} \), supported by a rigid, adjustable mount that enabled fine vertical positioning. Yaw and pitch were minimized to maintain normal alignment with the test cell. Fill-level measurements were tracked using two reference methods: an external scale visible in the camera’s field of view and an immersed probe used before each test. The fill level measurement uncertainty is estimated to be $\delta H_l = \pm \SI{5.0}{\mm}$.

The sloshing cell was mounted on aluminum feet via flanges attached to its end domes and securely bolted to a base plate (see bottom of Fig.~\ref{fig:experimental_setup}). This base plate was connected to two three-axis load cells (ME K3D120, $\pm\SI{1000}{\newton}$) for force measurements. The load cells were, in turn, coupled to a stainless steel fastening plate (excluded from Fig.~\ref{fig:experimental_setup}), which was mounted on the SHAKESPEARE sloshing table. This three-axis table enables controlled excitations along each axis, with amplitudes up to \( A_f = \pm\SI{45}{\milli\meter} \), forcing frequencies in the range $f_f  \in [0, 10]\,\si{\Hz}$ or \( \omega_f \in [0, 63]\,\si{\radian\per\s} \), and accelerations reaching up to \( A_f\omega_f^2/g \in [0, 1] \). Dynamic conditions are tracked through a triaxial accelerometer (Endevco Model 7298-2) with the uncertainty identified to be $\delta a/g = \pm 0.01$, and an optical displacement sensor (ODS) (Optoepsilon optoNCDT ILD1302-100) with $\delta A = \pm\SI{0.5}{\milli\meter}$.



\section{Data-Driven Identification of Sloshing Regimes} \label{sec3-regime_id}

This section presents the data-driven, image-based methodology developed to identify and classify sloshing regimes. An overview of the complete workflow is shown in Fig.~\ref{fig:flowchart_methodology}, which is organized into four main steps.

\textbf{Step 1: Image pre-processing.}  
Raw high-speed image sequences are stabilized and corrected to remove camera vibrations, small frame rotations, and non-uniform illumination. Keypoints are tracked between consecutive frames using the ORB (Oriented FAST and Rotated BRIEF) detector, and affine transformations are applied to align the sequence with a fixed reference frame. Residual tilt is corrected through Hough-line detection, while global intensity variations are compensated by histogram equalization and local mean subtraction. These operations, implemented with standard OpenCV routines~\citep{opencv_library}, ensure that only the fluid motion contributes to the subsequent analysis.

\textbf{Step 2: Multiscale modal decomposition.}  
The stabilized sequences are analyzed using the multiscale Proper Orthogonal Decomposition (mPOD, Sec.~\ref{sec:mPOD}), which constrains the energy-optimal POD with a filter bank to obtain spectrally separated temporal modes. This step provides a compact, frequency-resolved description of the sloshing dynamics suitable for both modal identification and reduced-order representation.

\textbf{Step 3: Prototype-based clustering.}  
Representative prototype cases are selected to define a reference modal library (Sec.~\ref{sec:labeling_data}).  
Each test case is projected onto this basis to quantify its similarity to the prototypes, yielding energy-based feature vectors that are subsequently clustered using unsupervised \(k\)-means. The prototypes are iteratively updated to the cluster medoids until consistent labels are obtained.

\textbf{Step 4: Classification and regime mapping.}  
A nonlinear Support Vector Machine (SVM) with a polynomial kernel (Sec.~\ref{sec:SVM}) is trained on the clustered data to delineate regime boundaries in the non-dimensional parameter space \((A_f/R,\;\omega_f/(2\,\omega_{1,0}))\). The classifier is informed by the Mathieu-equation stability analysis (Sec.~\ref{sec:Mathieu_eq}), linking experimental observations to the theoretical instability regions.

\subsection{Natural Frequency and Damping Ratio Identification} \label{sec:methodology_nat_freq}

In the absence of forcing, Eq.~\eqref{eq:mathieu_damped_d_dt} reduces the time evolution of each mode to that of a second-order damped oscillator~\citep{ibrahim_2005,abramson_1981}. Accordingly, the natural frequency of the antisymmetric longitudinal mode $(m,n) = (1,0)$, $\omega_{1,0}$, and its associated damping ratio, $\zeta_{1,0}$, were identified from free-decay tests. Alternative approaches are discussed by \citet{casiano2016damping}.

The experimental procedure consisted of first forcing the system near the $(1,0)$ mode resonance frequency until a fully developed periodic response was established, after which the excitation was stopped and the ensuing free decay was recorded. The forcing duration was set to \SI{60}{\s}, which was largely sufficient to ensure full development in all the investigated test cases. The forcing frequency for this initial phase was set to the theoretical value predicted by potential-flow analysis for a flat-ended cylindrical tank~\citep{hasheminejad_analytical_2017}. Throughout all experiments, the identification procedure relied on synchronized high-speed imaging and load-cell measurements providing the vertical forces $F_{z_1}(t)$ and $F_{z_2}(t)$ at the tank supports, together with the platform acceleration signal $a_z(t)$. These measurements were used to reconstruct the time evolution of the free-surface displacement and to verify the consistency of the hydrodynamic response.

Rather than applying conventional image-based tracking of the free surface --- which can suffer from spurious detection and case-specific parameter tuning~\citep{marques_experimental_2023,monteiro_microgravity,backlight_sloshing} --- the high-speed image sequences were analyzed using the mPOD approach, described in detail in the following Subsection. Balancing energy optimality and frequency selectivity, the temporal structure of the leading mPOD mode $\boldsymbol{\psi}_1$ (see Eq.~\eqref{eq:modal_decomposition}) accurately captures the dominant sloshing dynamics, while its associated spatial mode $\boldsymbol{\phi}_1$ confirms that the identified response corresponds to the expected $(m,n)=(1,0)$ antisymmetric mode.

The time series of all measured quantities --- $\psi_1(t)$, $F_{z_1}(t)$, $F_{z_2}(t)$, and $a_z(t)$ --- were analyzed using a continuous wavelet transform (CWT) to resolve the time-dependent evolution of their spectral content. The CWT provides a localized time–frequency representation, allowing the transient shift from the forced oscillation to the free-decay regime to be clearly identified. A complex Morlet wavelet was employed, with a bandwidth parameter of $6.0$, center frequency of $2.0$, and 64 voices per octave, providing a suitable balance between temporal and spectral resolution. The instantaneous dominant frequency is extracted as

\begin{equation}
f_r(t) = \operatorname*{arg\,max}_{f \in \mathcal{B}} P(t,f),
\end{equation} where $P(t,f)$ denotes the wavelet power, defined as the squared magnitude of the complex CWT coefficients. The frequency domain $\mathcal{B}$ is determined by the sampling frequency and the chosen wavelet parameters, and in this analysis, it was restricted to the band encompassing the fundamental sloshing mode. This approach provides a smooth estimate of the temporal evolution of the resonance frequency during the transition from the forced to the free-decay regime.

Once the external excitation is stopped, the dominant frequency $f_r(t)$ transitions from the forcing frequency toward the damped natural frequency of the $(1,0)$ mode (see Sec.~\ref{sec:nat_freq_experimental}). This behavior is clearly observed in the CWT ridge of the image-derived signal $\psi_1(t)$ and in the load-cell forces $F_{z_1}(t)$ and $F_{z_2}(t)$, whereas the accelerometer signal $a_z(t)$ primarily reflects the imposed platform motion and therefore captures only the pre–ring-down stage. 

The ring-down interval $t \in [t_1,t_2]$ is defined as a portion of the signal following the cessation of forcing. The initial time $t_1$ is selected when the subharmonic ridge power, $P_r(t)=|W(t,f_r(t))|^2$, first exceeds the residual power associated with the forcing component in $F_{z_1}$ and $F_{z_2}$. Within this window, the liquid oscillates freely at the damped natural frequency,
\begin{equation}
\omega_{d_{1,0}} = \omega_{1,0}\sqrt{1-\zeta_{1,0}^2}, 
\qquad
f_{1,0} = \frac{\omega_{1,0}}{2\pi},
\label{eq:damped_nat_freq}
\end{equation} which is estimated directly from the ridge as
\begin{equation}
f_{d_{1,0}} = \operatorname*{median}_{t\in[t_1,t_2]} f_r(t),
\qquad
\omega_{d_{1,0}} = 2\pi f_{d_{1,0}},
\label{eq:fd}
\end{equation} where the median acts as a robust estimator against local fluctuations. The damping ratio is obtained from the temporal decay of the ridge power within the same interval. Because the CWT ridge behaves as a time-localized band-pass filter centered on the mode of interest, the ridge power decays approximately exponentially as $P_r(t)\propto e^{-2\,\zeta_{m,n}\,\omega_{d_{m,n}}\,t}$ for a correctly isolated second-order response~\citep{tiziano_damping_wavelet, slavic_pendulum_damping, hans_civil_damping}. Taking the logarithm gives
\begin{equation}
\ln P_r(t) = -2\,\zeta_{m,n}\,\omega_{d_{m,n}}\,t + \ln P_0,
\label{eq:zeta_log_slop}
\end{equation} so that the slope of a first-order least-squares fit of $\ln P_r(t)$ versus $t$ over $[t_1,t_2]$ yields the mean damping ratio $\zeta_{m,n}$. The resulting values of $\omega_{1,0}$ and $\zeta_{1,0}$ serve as inputs for the Floquet analysis and are subsequently used to normalize the excitation parameters in the sloshing-regime maps of Sec.~\ref{sec:sloshing_maps}.

\subsection{Regime Prototype Definition via mPOD} \label{sec:mPOD}

The Multiscale Proper Orthogonal Decomposition (mPOD)~\citep{MENDEZ_2018256,Mendez_2020,MENDEZ_Poletti2024} is a data-driven modal decomposition that constrains the energy optimality of the Proper Orthogonal Decomposition (POD) with a filter bank that allows for the spectral separation of its modes. This combination isolates nearly harmonic dynamics undergoing transient evolution and it is thus well suited to the narrow-band yet nonstationary nature of sloshing responses. The mPOD of the video sequence provides a reduced-order space to encode the sloshing response from the high-speed videos (Step 2; see Fig. \ref{fig:flowchart_methodology}).

As in all data-driven decompositions, the grayscale image sequence acquired during the experiments is arranged into the snapshot matrix $\boldsymbol{D}\in\mathbb{R}^{n_s\times n_t}$, whose columns correspond to vectorized frames:
\begin{equation}
\boldsymbol{D} =
\begin{bmatrix}
g_1[1,1] & \cdots & g_{n_t}[1,1] \\
\vdots & \ddots & \vdots \\
g_1[n_x,1] & \cdots & g_{n_t}[n_x,1] \\
g_1[1,2] & \cdots & g_{n_t}[1,2] \\
\vdots & \ddots & \vdots \\
g_1[n_x,n_z] & \cdots & g_{n_t}[n_x,n_z]
\end{bmatrix}
\in \mathbb{R}^{n_s\times n_t},
\end{equation} where each image $g_k[i,j]\in[0,255]$ has spatial dimensions $n_s=n_x\times n_z$ and $n_t$ denotes the total number of frames. As in classical POD, the mPOD represents the data as
\begin{equation}
\boldsymbol{D} = \sum_{r=1}^{n_R}\sigma_r\,\boldsymbol{\phi}_r\,\boldsymbol{\psi}_r^{\top}
               = \boldsymbol{\Phi}\,\boldsymbol{\Sigma}\,\boldsymbol{\Psi}^{\top},
\label{eq:modal_decomposition}
\end{equation} where $\boldsymbol{\phi}_r$ and $\boldsymbol{\psi}_r$ are the spatial and temporal modes, and $\sigma_r$ is the associated amplitude.

In contrast with the classical POD, whose temporal modes are the eigenvectors of the correlation matrix 
$\boldsymbol{K}=\boldsymbol{D}^\top\boldsymbol{D}$ and may exhibit broad spectral content, the mPOD introduces a multiresolution analysis of $\boldsymbol{K}$.  
The correlation matrix is decomposed into band-limited components associated with $n_M-1$ non-overlapping frequency bands,
\begin{equation}
\boldsymbol{K}\approx \sum_{m=0}^{n_M-1} \boldsymbol{K}^{(m)},
\end{equation}
each obtained by filtering $\boldsymbol{K}$ in the frequency domain as
\begin{equation}
\boldsymbol{K}^{(m)} =
\boldsymbol{\Psi}_F
\!\left[
\left(\boldsymbol{\Psi}_F^{\dagger}\boldsymbol{K}\boldsymbol{\Psi}_F\right)
\odot
\left(\mathcal{H}_m^{\top}\mathcal{H}_m\right)
\right]
\boldsymbol{\Psi}_F^{\dagger},
\label{eq:km_expanded_correct}
\end{equation}
where $\boldsymbol{\Psi}_F\in\mathbb{C}^{n_t\times n_t}$ is the discrete Fourier transform matrix, $\mathcal{H}_m\in \mathbb{C}^{1\times n_t}$ collects the discrete transfer function of the $m$-th band-pass filter, and $\odot$ denotes the Hadamard product. The product $\mathcal{H}_m^\top \mathcal{H}_m$ ensures that $\boldsymbol{K}^{(m)}$ remains Hermitian and thus has real, orthogonal eigenvectors. Ensuring that the filter bank satisfies a partition of unity,  

\begin{equation}
\sum_{m=0}^{n_M-1} \mathcal{H}_m = \mathbf{1},
\qquad
\mathcal{H}_m \odot \mathcal{H}_{m'} = \mathbf{0}
\quad (m \neq m'),
\label{eq:partition_unity}
\end{equation} the spectral bands are non-overlapping and complete, guaranteeing that the temporal modes obtained from the eigenproblems
\begin{equation}
\boldsymbol{K}^{(m)}\,\boldsymbol{\psi}^{(m)}=\lambda^{(m)}\,\boldsymbol{\psi}^{(m)}
\label{eq:Km_eig}
\end{equation} are mutually orthogonal across scales. Aggregating the eigenvectors of all bands yields the full mPOD temporal basis $\boldsymbol{\Psi}$, which remains orthogonal while offering controlled spectral separation among its modes.

Once the temporal structures are computed, the spatial structures and modal amplitudes are recovered through temporal projection, following the classical snapshot-based POD framework~\citep{Mendez2023}:
\begin{equation}
\boldsymbol{\Phi} = \boldsymbol{D}\,\boldsymbol{\Psi}\,\boldsymbol{\Sigma}^{-1},
\qquad
\sigma_r = \bigl\|\boldsymbol{D}\,\boldsymbol{\psi}_r\bigr\|_2.
\label{eq:phi}
\end{equation}

A consequence of the spectral separation achieved by mPOD is that the resulting spatial structures are no longer strictly orthogonal, in contrast to those obtained via standard POD.  
Nevertheless, the method yields modes that are energetically optimal within each spectral band and temporally well localized, enabling a clear association between individual modes and specific sloshing frequencies. In this study, the leading temporal coefficient $\boldsymbol{\psi}_1$ captures the time-resolved evolution of the dominant sloshing mode, while the reduced modal space spanned by the leading mPOD modes provides a compact and physically interpretable representation in which distinct sloshing regimes can be clearly separated despite the transient dynamics.

\subsection{Data labeling and clustering} \label{sec:labeling_data}

The proposed data-labeling and clustering strategy follows a prototype-based, semi-supervised approach to group experimental realizations (videos) that exhibit similar flow dynamics or regime behavior, as encoded in their reduced feature representations (energy signatures derived from mPOD projections; see Step 3 in Fig. \ref{fig:flowchart_methodology}). The method begins with a set of representative prototypes \(p_j\), with \( j \in \{1,2,\ldots,n_P\}\), each corresponding to a distinct sloshing regime. The prototypes were selected from experimental realizations that best exemplify the characteristic dynamics of each regime. This selection was first carried out manually and then refined iteratively. For every prototype, the leading spatial modes obtained from its mPOD decomposition are retained to encode its dominant flow features in a reduced form. The collection of these mode sets defines a global reference basis onto which all other cases are projected for subsequent clustering and classification.

Denoting by \(m_j\) the number of mPOD modes retained for prototype \(p_j\), the global reference basis matrix is constructed as
\begin{equation}
\boldsymbol{\Phi}_{ref} =
\begin{bmatrix}
\boldsymbol{\phi}_{1}^{(p_1)} \;\;
\ldots \;\;
\boldsymbol{\phi}_{m_1}^{(p_1)} \;\;
\ldots \;\;
\boldsymbol{\phi}_{1}^{(p_{n_P})} \;\;
\ldots \;\;
\boldsymbol{\phi}_{m_{n_P}}^{(p_{n_P})}
\end{bmatrix},
\label{eq:reference_basis}
\end{equation} with \(\boldsymbol{\Phi}_{ref}\in\mathbb{R}^{n_s\times n_b}\) and \(n_b=\sum_{j=1}^{n_P} m_j\) the total number of basis vectors across all prototypes.   

Given this reference basis, the snapshot matrix \mbox{\(\boldsymbol{D}\in\mathbb{R}^{n_s\times n_t}\)} of any experimental realization --- whether part of the training set or unseen --- can be projected onto the reduced subspace through a standard linear projection:
\begin{equation}
\boldsymbol{A}_{proj} =
\left(\boldsymbol{\Phi}_{ref}^{\top}\boldsymbol{\Phi}_{ref}\right)^{-1}
\boldsymbol{\Phi}_{ref}^{\top}\boldsymbol{D}
\in\mathbb{R}^{n_b\times n_t}.
\label{eq:projection_A_proj}
\end{equation} The resulting coefficient matrix \(\boldsymbol{A}_{proj}\) provides a compact, physically interpretable representation of each realization in terms of its similarity to the prototype modes.

This operation linearly maps the videos $\boldsymbol{D}\in\mathbb{R}^{n_s\times n_t}$ into a set of $n_b$ time series, stored in $\boldsymbol{A}_{proj}\in\mathbb{R}^{n_b\times n_t}$, with $n_b\ll n_t$, that quantify the temporal activation of each reference mode in \eqref{eq:reference_basis}. This is thus a compression $n_s\rightarrow n_b$ of each image in the video. 

The following step consists of compressing the temporal dimension to $n_t\rightarrow 1$, which transforms each video into a compact feature vector $\boldsymbol{z}_i\in\mathbb{R}^{n_b}$. This is carried out by first filtering the time series in $\boldsymbol{A}_{proj}$ using the same multi-resolution filter bank employed in the mPOD analysis, and then computing the $\ell_2$-norm ($\left\| ^{.} \right\|_2$) of each filtered time series.

Let \(\boldsymbol{h}_m = [\,h_m[0], \,h_m[1], \ldots,\, h_m[n_O - 1]\,]^\top \in \mathbb{R}^{n_O}\) denote the impulse response of the FIR filter associated with scale \( m \), where \( \tau = 0, 1, \ldots, n_O - 1 \) is the discrete time index and \( n_O \) is the filter order. The \( i \)-th time series \(\boldsymbol{a}_i^{(p_j)} \in \mathbb{R}^{n_t}\), extracted from \(\boldsymbol{A}_{proj}\), is then filtered using a zero-phase FIR operation. The filtered signal is evaluated at the discrete time index \( k \in \{0, 1, \ldots, n_t - 1\} \) and given by:
\begin{align}
a_{i,filt}^{(p_j)}[k]
&=
\sum_{n=0}^{n_O - 1} h_m[n]
\Big(
\sum_{\tau=0}^{n_O - 1} h_m[\tau]\,
a_{i}^{(p_j)}[k + n - \tau]
\Big).
\end{align} Here \(\boldsymbol{a}_i^{(p_j)}\) is obtained through the projection on the prototype basis element \(\boldsymbol{\phi}_{i}^{(p_j)}\), and \(a_{i,filt}^{(p_j)}[k]\) is its filtered counterpart. The feature vector is then computed as:
\begin{align}
\boldsymbol{z}_i = &
\left[
\left\| \boldsymbol{a}_{1,filt}^{(p_1)} \right\|_2,
\ldots,
\left\| \boldsymbol{a}_{m_1,filt}^{(p_1)} \right\|_2,
\right. \nonumber \\
& \left.
\ldots \,\,\,
\left\| \boldsymbol{a}_{1,filt}^{(p_{n_P})} \right\|_2,
\ldots,
\left\| \boldsymbol{a}_{m_{n_P},filt}^{(p_{n_P})} \right\|_2
\right]^\top
\in \mathbb{R}^{n_b}.
\label{eq:feature_vector}
\end{align} The entries of this vector collect the similarity scores associated with all mPOD modes across all prototypes. To quantify the contribution of a specific prototype \(p_j\) to a given test case, the scores corresponding to its modes are grouped. Denoting by \(\mathcal{I}_p\) the set of indices in \(\boldsymbol{z}_i\) associated with prototype \(p_j\), the prototype score is defined as
\begin{equation}
s_i^{(p_j)} = \left\| (\boldsymbol{z}_i)_{\mathcal{I}_p} \right\|_2,
\qquad
\tilde{s}_i^{(p_j)} = 
\frac{s_i^{(p_j)}}{\sum_{j=1}^{n_P} s_i^{(p_j)}}.
\label{eq:prototype_presence}
\end{equation}

Feature vectors defined in \eqref{eq:feature_vector}, along with the prototype presence metrics in \eqref{eq:prototype_presence}, are computed for all \( i = 1, \ldots, N_{pts} \) videos. Then, to identify the underlying sloshing regimes, an unsupervised clustering is performed in this reduced-order space using standard \( k \)-means clustering \citep{Ikotun2023}. This algorithm partitions the set of feature vectors \( \{ \boldsymbol{z}_i \}_{i=1}^{N_{pts}} \) into \( K \) disjoint clusters \( \{ \mathcal{C}_0, \dots, \mathcal{C}_{K-1} \} \) by minimizing the within-cluster variance, i.e., the distance between each point and its assigned cluster centroid:
\begin{equation}
\min_{\{ \mathcal{C}_0, \dots, \mathcal{C}_{K-1} \}} \sum_{c=1}^K \sum_{\boldsymbol{z}_i \in \mathcal{C}_c} \left\| \boldsymbol{z}_i - \boldsymbol{\mu}_c \right\|_2^2,
\label{eq:kmeans_objective} 
\end{equation} where \( \boldsymbol{\mu}_c \) denotes the centroid of cluster \( \mathcal{C}_c \). At each iteration, the centroid is updated by averaging the feature vectors belonging to the cluster:
\begin{equation}
\boldsymbol{\mu}_c = \frac{1}{|\mathcal{C}_c|} \sum_{\boldsymbol{z}_i \in \mathcal{C}_c} \boldsymbol{z}_i,
\label{eq:kmeans_centroid}
\end{equation} and the feature vector \( \boldsymbol{z}_i \) is then reassigned to the cluster whose centroid is closest in the Euclidean sense:
\begin{equation}
\boldsymbol{z}_i \in \mathcal{C}_c \quad \text{if} \quad \left\| \boldsymbol{z}_i - \boldsymbol{\mu}_c \right\|_2^2 \leq \left\| \boldsymbol{z}_i - \boldsymbol{\mu}_j \right\|_2^2, \quad \forall j = 1, \dots, K.
\label{eq:kmeans_assignment}
\end{equation}


To ensure that the final clustering is independent of the initial prototype selection, the entire procedure --— from the prototype basis construction \eqref{eq:reference_basis} to the clustering and score assignment \eqref{eq:kmeans_assignment} --- is repeated iteratively. At each iteration, the new prototypes are defined as the medoids of the clusters, i.e., the realizations minimizing the intra-cluster Euclidean distance in the reduced feature space. The iterative process continues until convergence, defined as either (i) an Adjusted Rand Index (ARI) between consecutive clusterings exceeding a prescribed threshold (here ARI~$\geq~0.98$)~\citep{ARI_article}, or (ii) no change in the medoid set relative to the previous iteration (prototype overlap). The final cluster labels obtained upon convergence are then used in the classification stage.

\subsection{Regime Classification via physics informed SVM} \label{sec:SVM}

The proposed regime-classification strategy employs a physics-informed Support Vector Machine (SVM) to delineate the boundaries between the sloshing regimes identified through clustering, as shown in Step~4 of Fig.~\ref{fig:flowchart_methodology}.  
The SVM~\citep{Cortes1995} provides a supervised mapping from the labeled experimental realizations onto the nondimensional parameter space:
\begin{equation}
\boldsymbol{x}_i = \begin{bmatrix} A_f / R \\[4pt] \omega_f / (2 \, \omega_{1,0})\ \end{bmatrix} \in \mathbb{R}^2,
\label{eq:parametric_space}
\end{equation} enabling prediction of the expected regime for new operating conditions. Physical consistency is introduced through a weighting scheme that incorporates analytical stability information from the Mathieu-equation Floquet analysis, yielding a hybrid framework that blends data-driven classification with the governing physics of parametric resonance (see Subsec. \ref{sec:sloshing_maps}).

The experimental campaign yields a dataset of \( N_{pts} \) samples: \(\boldsymbol{x}_1, \ldots, \boldsymbol{x}_{N_{pts}}\). For each sample, the prototype-based clustering method of Sec.~\ref{sec:labeling_data} assigns a label \( \mathcal{C}_0,\ldots,\mathcal{C}_{K-1} \) corresponding to a distinct sloshing regime.  
The objective of the classification step is to construct a continuous decision function $f(\boldsymbol{x}):\boldsymbol{x}\!\rightarrow\!\{\mathcal{C}_0,\ldots,\mathcal{C}_{K-1}\}$ that generalizes to unseen data and provides smooth decision boundaries between regimes.  
Following the one-vs-one (OvO) strategy~\citep{crammer2001algorithmic,hsu2002comparison} adopted here, \(K(K{-}1)/2\) binary SVM classifiers are trained, each separating a unique pair of classes. During inference, their pairwise decision functions are aggregated into per-class scores \(\boldsymbol{F}_i = [\,f_{i0},\,f_{i1},\ldots,f_{i,K-1}\,]\), where each \(f_{ik}\) encodes the combined vote and margin confidence for class \(k\).

Each binary classifier is defined by a kernel-based decision function
\begin{equation}
f(\boldsymbol{x}) = \sum_{i=1}^{N_s} \alpha_i y_i\, 
\mathcal{K}(\boldsymbol{x}_i,\boldsymbol{x}) + b,
\label{class_Func}
\end{equation} where \(N_s\) denotes the number of support vectors (i.e., the training samples with \(\alpha_i > 0\) after solving the dual problem in \eqref{eq:svm_dual_weighted_a} and \eqref{eq:svm_dual_weighted_b}), \(\alpha_i\) and \(b\) are learned coefficients, \(y_i \in \{-1,+1\}\) are binary class labels, \(\boldsymbol{x}_i\) are the corresponding support vectors, and \(\mathcal{K}(\boldsymbol{x},\boldsymbol{x}')\) is a kernel function measuring similarity in the input space. A third-order polynomial kernel is used:
\begin{equation}
\mathcal{K}(\boldsymbol{x},\boldsymbol{x}') = \bigl(\gamma\,\boldsymbol{x}^\top\boldsymbol{x}' + c\bigr)^3,
\end{equation} with $\gamma$ controlling the influence of individual training points and $c$ the bias term. The hyperparameters are set to \(\gamma = 1/(2\,\operatorname{var}(\boldsymbol{x}))\) and \(c=5\).

The classification uncertainty is then quantified through a top-margin-based criterion that compares the two highest decision scores in $\boldsymbol{F}_i$. Sorting the scores as $f_{i(1)} \ge f_{i(2)} \ge \dots \ge f_{i(K)}$ the confidence gap is defined as: 
\begin{equation}
g_i = f_{i(1)} - f_{i(2)},  \quad g_i \ge 0,
\end{equation} which quantifies the separation between the most probable class and its nearest competitor. Across the parametric space, the gap is normalized to \([0,1]\) and inverted to yield the uncertainty measure
\begin{equation}
u_i = 1 - \frac{g_i - g_{\min}}{g_{\max} - g_{\min}},
\end{equation} with \( g_{\min} \) and \( g_{\max} \) denoting the minimum and maximum gap values. Thus, large \(u_i\) values mark regions of low classifier confidence, typically near decision boundaries. A quantile threshold of \SI{2.5}{\%} is applied, retaining only the uppermost uncertain points to delineate these regions.

The Support Vector Machine used here follows the weighted formulation implemented in \texttt{scikit-learn}. For a given OvO classifier, let \(N\) denote the number of training samples belonging to the two classes being separated. The dual optimization problem solves for the multipliers 
\(\boldsymbol{\alpha} = [\alpha_1,\ldots,\alpha_N]^\top\) and the bias \(b\) via:
\begin{align}
\min_{\boldsymbol{\alpha}} \quad &
\frac{1}{2}\,\boldsymbol{\alpha}^\top \mathbf{Q}\,\boldsymbol{\alpha}
- \mathbf{1}^\top \boldsymbol{\alpha},
\label{eq:svm_dual_weighted_a} \\[3pt]
\text{s.t.} \quad &
0 \le \alpha_i \le C\,w_i, \qquad
\sum_{i=1}^{N} y_i\,\alpha_i = 0,
\label{eq:svm_dual_weighted_b}
\end{align} where \(Q_{ij} = y_i y_j\,\mathcal{K}(\boldsymbol{x}_i,\boldsymbol{x}_j)\) for \(i,j = 1,\ldots,N\), \(C > 0\) is the penalty parameter, and \(w_i\) are sample-specific weights. After optimization, the subset of training samples with \(\alpha_i > 0\) constitute the support vectors in Eq.~\eqref{class_Func}. The bias \(b\) is recovered from the Karush–Kuhn–Tucker conditions.

To encode prior physical knowledge, the weights \(w_i\) are derived from the analytical neutral-stability boundaries of the inviscid Mathieu equation (Sec.~\ref{sec:Mathieu_eq}). Near these boundaries, the dynamics transition between stable and unstable behavior, increasing classification uncertainty. 

For each sample, the Euclidean distance \(d_i\) to the theoretical boundary is computed by orthogonal projection onto the boundary curve, and a normalized measure
\begin{equation}
t_i = \operatorname{clip}\!\left(
\frac{d_i-\delta_1}{\delta_2-\delta_1},\,0,\,1
\right)
\end{equation} is used to define individual weights:
\begin{equation}
w_i = w_{\min} + (1-w_{\min})\,t_i.
\end{equation} Here, \(\delta_1\) and \(\delta_2\) specify the inner and outer limits of a guard band, and \(w_{\min}\) sets the minimum confidence assigned to near-boundary samples. The operator \(\operatorname{clip}(x,a,b) = \min(\max(x,a),b)\) denotes elementwise bounding to the interval \([a,b]\).

Weights are then normalized and clipped to avoid extreme scaling:
\begin{equation}
w_i \leftarrow
\operatorname{clip}\!\left(
\frac{w_i}{\overline{w}},\,c_{\min},\,c_{\max}
\right),
\end{equation} where \(\overline{w}\) is the mean weight over the training set. With \(\delta_1 = 0.01\), \(\delta_2 = 0.10\), \(w_{\min} = 0.01\), and \((c_{\min},c_{\max}) = (0.1,5.0)\), this penalization scheme downweights samples close to the theoretical boundaries, guiding --- but not constraining --- the classifier toward physically consistent decision regions.

\section{Results}\label{sec4-results}

\begin{figure*}[t!]
\centering
\includegraphics[width=1.0\textwidth]{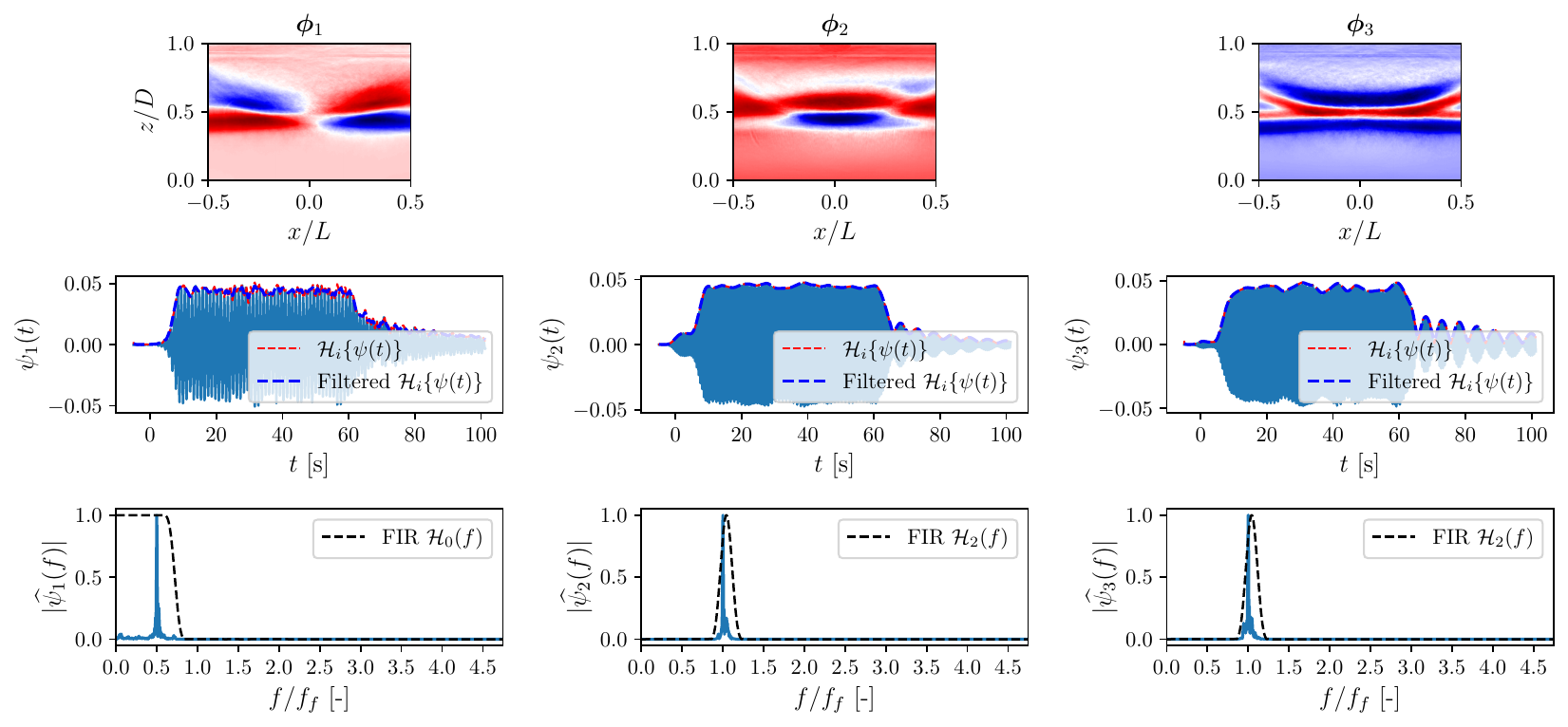}
\caption{mPOD analysis for $H_l/D = 0.50$ at the forcing condition $A_f = \SI{39.7}{\milli\meter}$ and $f_f = \SI{1.85}{\hertz}$, performed over the full image acquisition interval $t \in [-5.0,\,100.0]$ seconds. Spatial modes $\boldsymbol{\phi}_{r}$, temporal coefficients $\boldsymbol{\psi}_{r}$, and their normalized frequency spectra $|\widehat{\boldsymbol{\psi}}_{r}|$ are presented for the three leading modes. The sloshing signal starts at $t = \SI{0.0}{\s}$ and the dominant mode shape $\boldsymbol{\phi}_1$ has a single-node antisymmetric longitudinal structure with a subharmonic response.}
\label{fig:nat_freq_POD_modes}
\end{figure*}

The results are organized into four subsections, each analyzing a distinct stage of the proposed methodology (see Fig.~\ref{fig:flowchart_methodology}), leading to the final sloshing regime map. For clarity, Subsecs.~\ref{sec:prototype_library} -- \ref{sec:clustering_results} illustrate the complete workflow using the intermediate fill level $H_l/D = 0.50 \pm 0.04$, while Subsec.~\ref{sec:sloshing_maps} presents the final regime maps for all tested fill ratios $H_l/D \in [0.40; 0.67]$.

For the case $H_l/D = 0.50$, a total of 73 test points were investigated through 146 runs to ensure repeatability (see Table~\ref{tab:experimental_matrix}). The forcing amplitude spanned $A_f / R \in [0.02, 0.62]$ and $A_f \omega_f^2 / g \in \pm [0.02, 0.77]$, while the forcing frequency ratio covered $\omega_f / (2\,\omega_{1,0}) \in [0.90, 1.20]$. The natural frequency of the first longitudinal sloshing mode was identified as $\omega_{1,0} \approx \SI{6.08}{\radian\per\second}$ from free-decay tests, as detailed in Sec.~\ref{sec:nat_freq_experimental}.

\subsection{Natural Frequency and Damping Ratio} \label{sec:nat_freq_experimental}

\begin{figure*}[t]
  \centering
  \subfigure{%
    \includegraphics[width=0.49\textwidth]{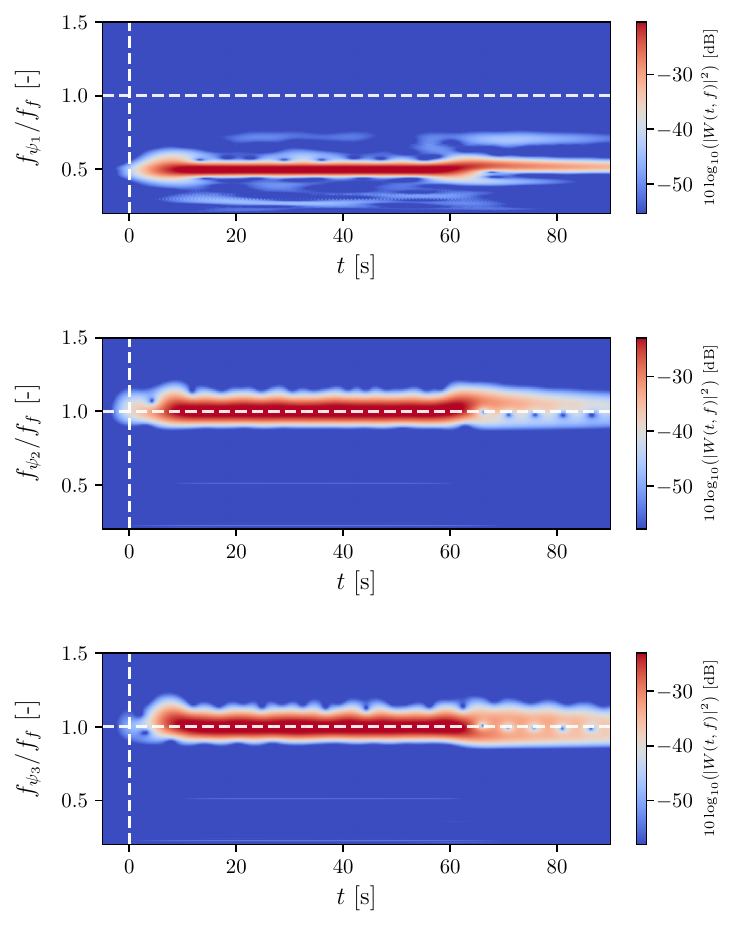}%
    \label{fig:nat_freq_wavelet-POD}}
  \hfill
  \subfigure{%
    \includegraphics[width=0.49\textwidth]{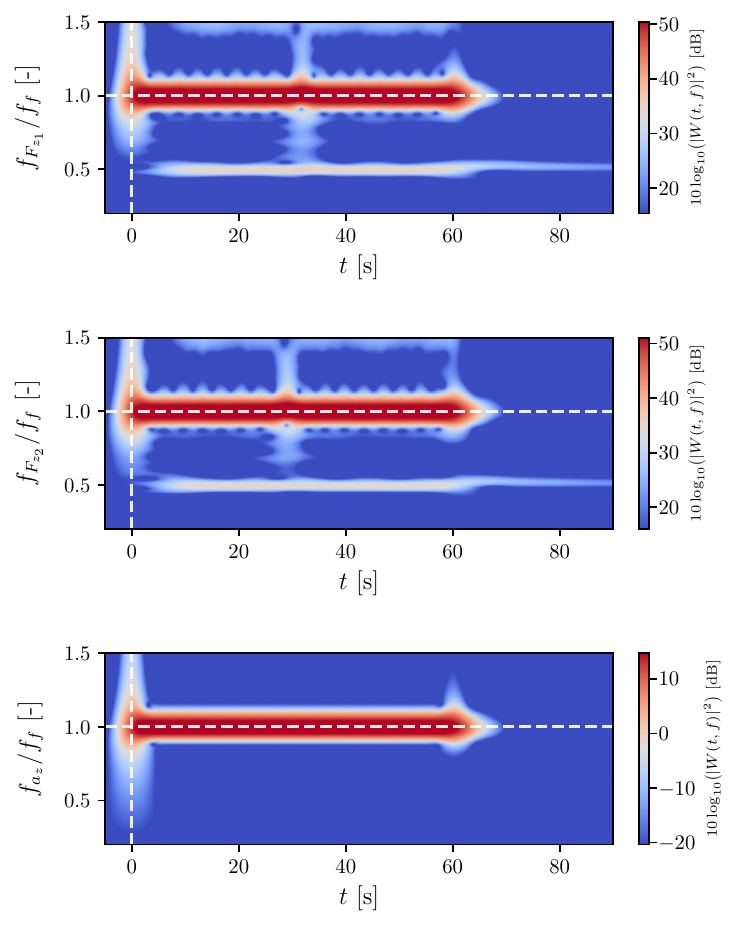}%
    \label{fig:nat_freq_wavelet-Sensors}}

\caption{Wavelet scalograms (time–frequency representation) obtained from the complex Morlet transform (bandwidth 6.0, center frequency 2.0) of the temporal coefficients of the three most energetic POD modes ($\boldsymbol{\psi}_1$, $\boldsymbol{\psi}_2$ and $\boldsymbol{\psi}_3$) in Fig. \ref{fig:nat_freq_POD_modes}, the two load-cell force signals ($F_{z_1}$ and $F_{z_2}$), and the accelerometer ($a_z$) over the interval $t \in [-5.0, 90.0]$ seconds. The sloshing signal is initiated at $t = \SI{0.0}{\second}$. The frequency axis is normalized by the forcing frequency $f_f = \SI{1.85}{\Hz}$, and wavelet power is represented on a logarithmic scale using $ P_{\log}(t,f) = 10\log_{10}(|W(t,f)|^2)$.}
\label{fig:nat_freq_wavelet}
\end{figure*}

The natural frequency $\omega_{1,0}$ and damping ratio $\zeta_{1,0}$ of the antisymmetric longitudinal mode $(m=1,\,n=0)$ were obtained from the free-decay tests introduced in Subsec.~\ref{sec:methodology_nat_freq}. Once the external excitation was removed, the tank–fluid system exhibited undriven oscillations dominated by a single mode, confirming that the extracted parameters correspond to a pure $(1,0)$ response rather than a multimodal superposition.

\begin{table*}[t!]
\caption{Statistical summary of the free–decay estimates of the fundamental antisymmetric sloshing mode \((m=1,\, n= 0)\) at \(H_l/D=0.50\). Reported are the forcing amplitude \(A_f\), forcing frequency \(f_f\), per test mean damped frequency \(\bar{f}_{d_{1,0}}\), damped angular frequency \(\bar{\omega}_{d_{1,0}}\), damping ratio \(\bar{\zeta}_{1,0}\), and the averaged natural frequency \(\bar{\omega}_{1,0}\) with a \SI{95}{\%} confidence interval based on \(N=4\) tests.}
\centering
\renewcommand{\arraystretch}{1.10}
\begin{tabular*}{\textwidth}{p{0.1\textwidth}@{\extracolsep{\fill}}p{0.1\textwidth}@{\extracolsep{\fill}}|cccc}
\toprule
$A_f$ [\unit{mm}] & $f_f$ [\unit{Hz}] & 
$\bar{f}_{d_{1,0}}$ [\unit{Hz}] &
$\bar{\omega}_{d_{1,0}}$ [\unit{\radian \per \s}] & 
$\bar{\zeta}_{1,0}$ [-]  & 
$\bar{\omega}_{1,0}$ [\unit{\radian \per \s}] \footnotesize(\SI{95}{\%} CI; $N=4$) \\
\midrule[1.25pt]
39.7 & 1.85 & 0.96 & 6.05 & 0.005 & \multirow{4}{*}[\dimexpr-0.3ex]{\(6.08 \pm 0.16 \)} \\
28.4 & 1.84 & 0.95 & 5.96 & 0.004 & \\ 
32.2 & 2.04 & 0.99 & 6.19 & 0.004 & \\ 
36.7 & 1.94 & 0.98 & 6.14 & 0.005 & \\ 
\bottomrule
\end{tabular*}
\label{tab:omega_n_stats_50}
\end{table*}

Fig.~\ref{fig:nat_freq_POD_modes} shows the three leading mPOD modes for $H_l/D = 0.50$ at $A_f = \SI{39.7}{\mm}$ and $f_f = \SI{1.85}{\Hz}$, including their spatial structures $\boldsymbol{\phi}_{r}$, temporal coefficients $\boldsymbol{\psi}_{r}$ with Hilbert envelopes $\mathcal{H}_i$, and frequency spectra $|\hat{\boldsymbol{\psi}}_r|$. The leading spatial mode $\boldsymbol{\phi}_1$ is antisymmetric, with a nodal line at $x/L=0$ and antinodes at $x/L=\pm0.5$, matching the expected sloshing eigenmode $(m,n)=(1,0)$ shape \citep{han_semi-analytical_2021}. During free decay, its spectrum (and the wavelet ridge) exhibits a dominant peak at the mode’s natural frequency $f_{1,0}$ or $\omega_{1,0}$.

After the onset of motion at $t = \SI{0.0}{\s}$, the temporal coefficient $\boldsymbol{\psi}_1$ grows exponentially and reaches a limit-cycle oscillation at $t \approx \SI{7.5}{\s}$. The spectrum shows a dominant frequency at $f \approx \SI{0.93}{\Hz}$, corresponding to the $1/2$-subharmonic of the driving frequency and confirming a Faraday-type instability with liquid oscillation at half the forcing frequency. Between $t = \SI{60.0}{\s}$ and $t = \SI{65.0}{\s}$, the imposed motion was exponentially attenuated to prevent excitation of higher-order modes, with the fully free decay starting at $t \sim \SI{65.0}{\s}$.

Fig.~\ref{fig:nat_freq_wavelet} displays the time-frequency representation of the three temporal coefficients ($\boldsymbol{\psi}_1$, $\boldsymbol{\psi}_2$ and $\boldsymbol{\psi}_3$) in Fig. \ref{fig:nat_freq_POD_modes}, the two load-cell vertical signals ($F_{z_1}$, $F_{z_2}$), and the vertical acceleration ($a_z$) over the interval $t \in [-5.0, 90.0]$ seconds. The vertical frequency axis is normalized by the forcing frequency $f_f$, emphasizing the drive and its (sub)harmonics. Power is shown in decibels [\unit{\decibel}] as $P_{\log}(t,f)~=~10\log_{10}\bigl(|W(t,f)|^2\bigr)$ where $W(t,f)$ denotes the CWT coefficient. 

In the force spectra ($f_{F_{z_1}}/f_f$ and $f_{F_{z_2}}/f_f$), a dominant ridge appears at the driving frequency while the excitation is active, due to the combined inertia of the liquid and tank. Once parametric sloshing develops, a weaker ridge emerges at $t \approx \SI{7.5}{\s}$ near $f_f/2$, consistent with the natural frequency $f_{1,0}$ and confirming the expected subharmonic response. A $3f_f/2$ sideband also appears, arising from nonlinear coupling between the drive and the subharmonic oscillation.

The accelerometer spectra ($f_{a_z}/f_f$) display only the primary forcing frequency, as expected for the rigid-body motion, while the subharmonic ridge persists in $F_{z_1}$, $F_{z_2}$, and $\boldsymbol{\psi}_1$, confirming the link between image-derived and measured quantities. Nonlinear coupling in $\boldsymbol{\psi}_1$ is also evidenced by mild spectral broadening and weak sidebands. At $t \sim \SI{65.0}{\s}$, a slight upward shift of the subharmonic ridge marks the transition to the free-decay phase. This ridge remains active at nearly constant frequency, while its power decays as viscous effects damp the motion. Since the damping is weak ($\zeta_{m,n} \ll 1$), the undamped and damped natural frequencies are nearly identical, $\omega_{m,n} \approx \omega_{d_{m,n}}$~\citep{ibrahim_2005}. For this test case, the subharmonic ridge is consistent across $\boldsymbol{\psi}_1$, $F_{z_1}$, and $F_{z_2}$, giving a mean damped frequency $\bar{f}_{d_{1,0}} \approx 0.525 f_f$ or $\bar{f}_{d_{1,0}} \approx \SI{0.97}{\Hz}$.

\begin{figure*}[t!]
  \centering
  \subfigure{%
    \includegraphics[width=1.0\textwidth]{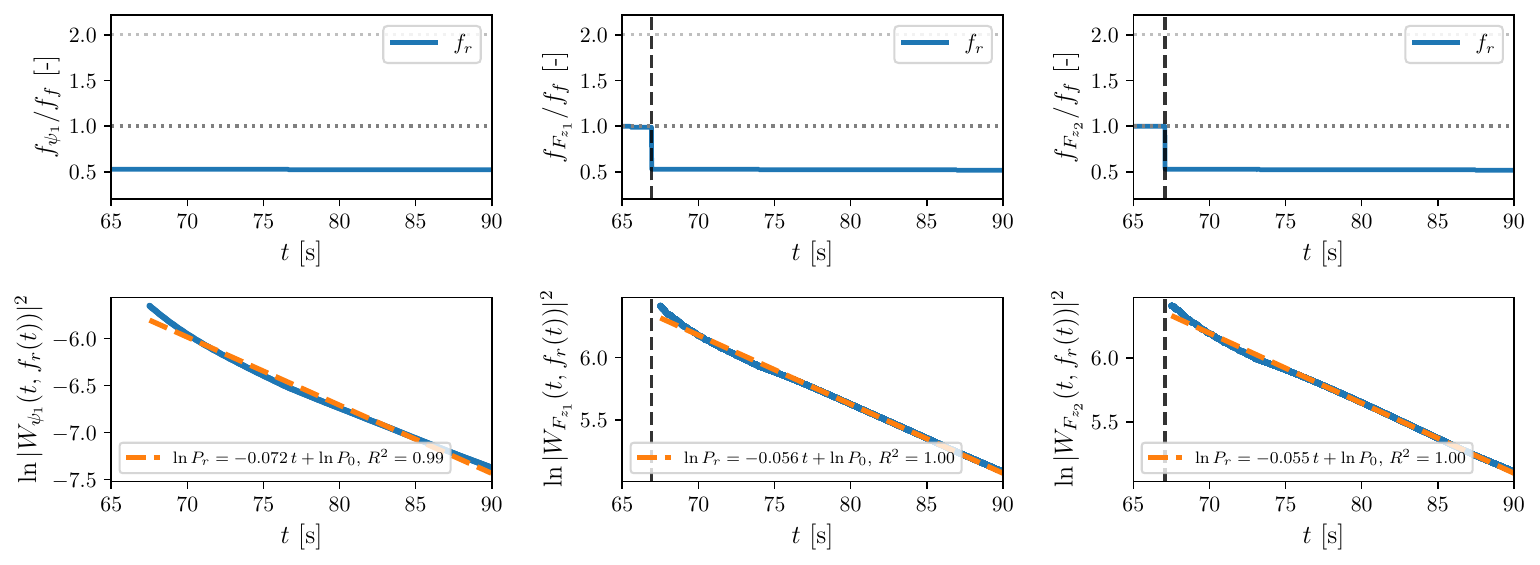}%
    }

  \caption{Wavelet ridge diagnostics for $H_l/D = 0.50$ at $A_f = \SI{39.7}{\milli\meter}$ and $f_f = \SI{1.85}{\hertz}$. Columns correspond to the three signals ($\boldsymbol{\psi}_1$, $F_{z_1}$, $F_{z_2}$). Top row: ridge frequency $f_r(t)/f_f$ identified as $\arg\max_{f\in\mathcal{B}}\,P(t,f)$; the dashed vertical lines indicate ridge jumps. Bottom row: logarithmic ridge power $\ln P_r(t)$ with least-squares fit $\ln P_r(t)=m_{\ln P_r}\,t+P_0$ over the ring-down window $[67.5, 90.0]$ (reported $m_{\ln P_r} = - 2\zeta_{1,0} \omega_{d_{1,0}}$ and $R^2$ shown in each panel).}

  \label{fig:nat_freq_wavelet_ridge}
\end{figure*}

\begin{figure}[b!]
\centering
\includegraphics[width=0.50\columnwidth]{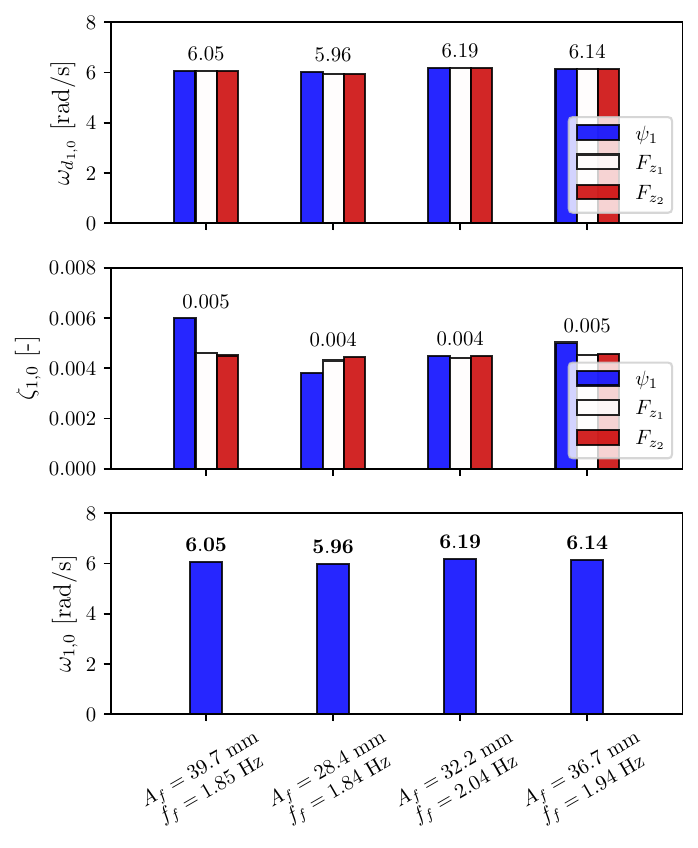}
\caption{Damped natural frequency $\omega_{d_{1,0}}$, damping ratio $\zeta_{1,0}$, and undamped natural frequency $\omega_{1,0}$ of the antisymmetric sloshing mode $(m=1,\,n=0)$, estimated from the leading mPOD coefficient $\boldsymbol{\psi}_1$ and the vertical forces $F_{z_1}$ and $F_{z_2}$. Results correspond to four free-decay tests with $A_f = \{39.7,\,28.4,\,32.2,\,36.7\}\,\mathrm{mm}$ and $f_f = \{1.85,\,1.84,\,2.04,\,1.94\}\,\mathrm{Hz}$. The subharmonic frequency is identified from the wavelet ridge, and $\zeta_{1,0}$ from the exponential decay of its power. Mean values for each test are shown as bar groups.}
\label{fig:nat_freq_bar_estimates}
\end{figure}

To achieve robust results the damped natural frequency extraction is repeated across three additional independent tests (following the methodology above), where $\boldsymbol{\phi}_1$ highlights the targeted mode \mbox{($m=1,\,n=0$)}; within each test, the three measurement methods (mPOD mode and the two load cells) are considered correlated and are therefore first averaged to provide a single per-test estimate. At $H_l/D = 0.50$, these additional test cases were performed at $A_f = \{28.4,\,32.2,\,36.7\}\,\unit{\mm}$ and $f_f = \{1.84,\,2.04, \, 1.94 \}\,\unit{\Hz}$, respectively, and are also used in the sloshing maps proposed in Sec. \ref{sec:sloshing_maps}. Statistical uncertainty on the mean was quantified with a two-sided \SI{95}{\%} Student’s $t$-interval based on the standard error $s/\sqrt{N}$, where $s$ is the unbiased standard deviation. This yields a conservative pooled estimate of $\bar{f}_{d_{1,0}} = 0.97 \pm \SI{0.03}{\Hz}$ (\SI{95}{\%} CI; $N=4$), such that $\bar{\omega}_{d_{1,0}} \approx \SI{6.08}{\radian \per \s}$. 
Analysis of the load-cell spectra during the free-decay period ($t > \SI{67.5}{\s}$) reveals a clear single ridge, while the temporal coefficient $\boldsymbol{\psi}_1$ shows slight amplitude modulation instead of a smooth exponential decay (see Figs.~\ref{fig:nat_freq_POD_modes} and~\ref{fig:nat_freq_wavelet-POD}). This modulation produces a weak beating pattern, making the raw $\boldsymbol{\psi}_1$ signal unsuitable for direct damping estimation using classical approaches such as the logarithmic decrement~\citep{dodge_2000, arndt_damping} or Hilbert-envelope methods~\citep{constantin_analysis_2021, constantin_nonlinear_2023} without prior band-pass filtering.

Fig.~\ref{fig:nat_freq_wavelet_ridge} summarizes the damping identification procedure applied to $\boldsymbol{\psi}_1$, $F_{z_1}$, and $F_{z_2}$. The top row shows the dominant wavelet ridge frequency, with a vertical marker indicating the instant when the subharmonic ridge power surpasses the forced response, marking the onset of free decay. For this case ($A_f = \SI{39.7}{\mm}$, $f_f = \SI{1.85}{\Hz}$), the transition occurs at $t \approx \SI{66}{\s}$, consistent with Fig.~\ref{fig:nat_freq_wavelet}, and the fitting window was set to $t_1 = \SI{67.5}{\s}$ and $t_2 = \SI{90.0}{\s}$. The bottom row presents the logarithmic ridge power $\ln P_r(t)$ and the corresponding linear least-squares fit. The high coefficients of determination ($R^2 \geq 0.99$) confirm that a single exponentially decaying mode was successfully isolated, with $\boldsymbol{\phi}_1$ validating the expected $(m,n)=(1,0)$ structure. 

Fig.~\ref{fig:nat_freq_bar_estimates} compiles the estimates of the damped and undamped natural frequencies and the damping ratios obtained across the four independent tests ($N=4$). The mean damping ratio is $\bar{\zeta}_{1,0} = 0.005 \pm 0.001$, determined using the same CWT-based ridge analysis employed for the damped frequency. The undamped natural frequency for each test was then computed as $\omega_{{1,0}_i} = \omega_{d_{{1,0}_i}} / \sqrt{1-\zeta_{{1,0}_i}^{2}}$, with $i = 1, \dots, 4$, and the average taken with a two-sided 95\% Student’s $t$-interval, yielding $\bar{\omega}_{1,0} = 6.08 \pm \SI{0.15}{\radian\per\s}$ (see Table~\ref{tab:omega_n_stats_50}). Consistent with previous studies~\citep{damping_1969_abramson, dodge_2000, arndt_damping}, this low-order sloshing mode exhibits a small damping ratio, validating the approximation $\bar{\omega}_{1,0} \approx \bar{\omega}_{d_{1,0}}$ for the present horizontal geometry.

\subsection{Prototype library} \label{sec:prototype_library} 

\begin{figure*}[htp!]
\centering
\subfigure[Prototype $p_1$: $A_f/R = 0.02, \,\, \omega_f/(2\,\omega_{1,0}) = 1.00$.]{
    \includegraphics[width=1\textwidth]{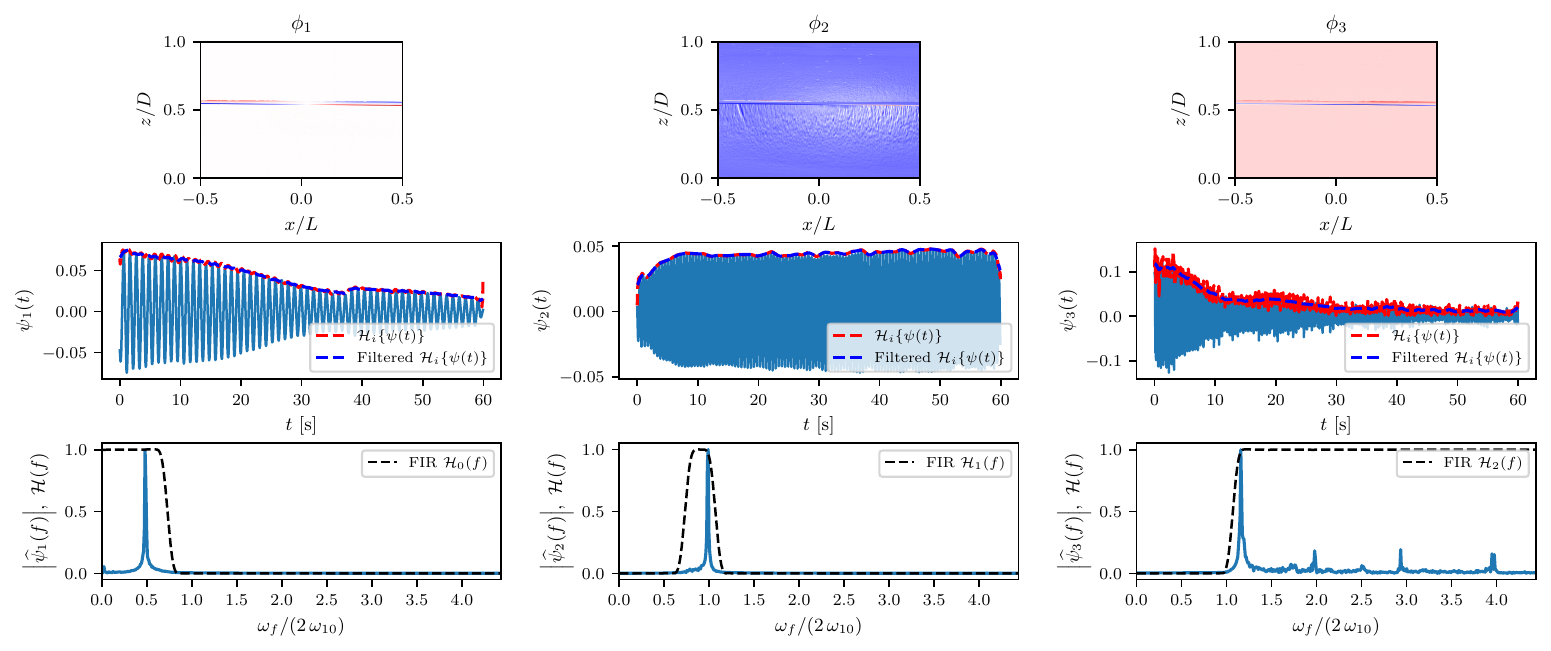}
    \label{fig:p1_stable_prototype}
}
\subfigure[Prototype $p_2$: $A_f/R = 0.13, \,\, \omega_f/(2\,\omega_{1,0}) = 1.00$.]{
    \includegraphics[width=1\textwidth]{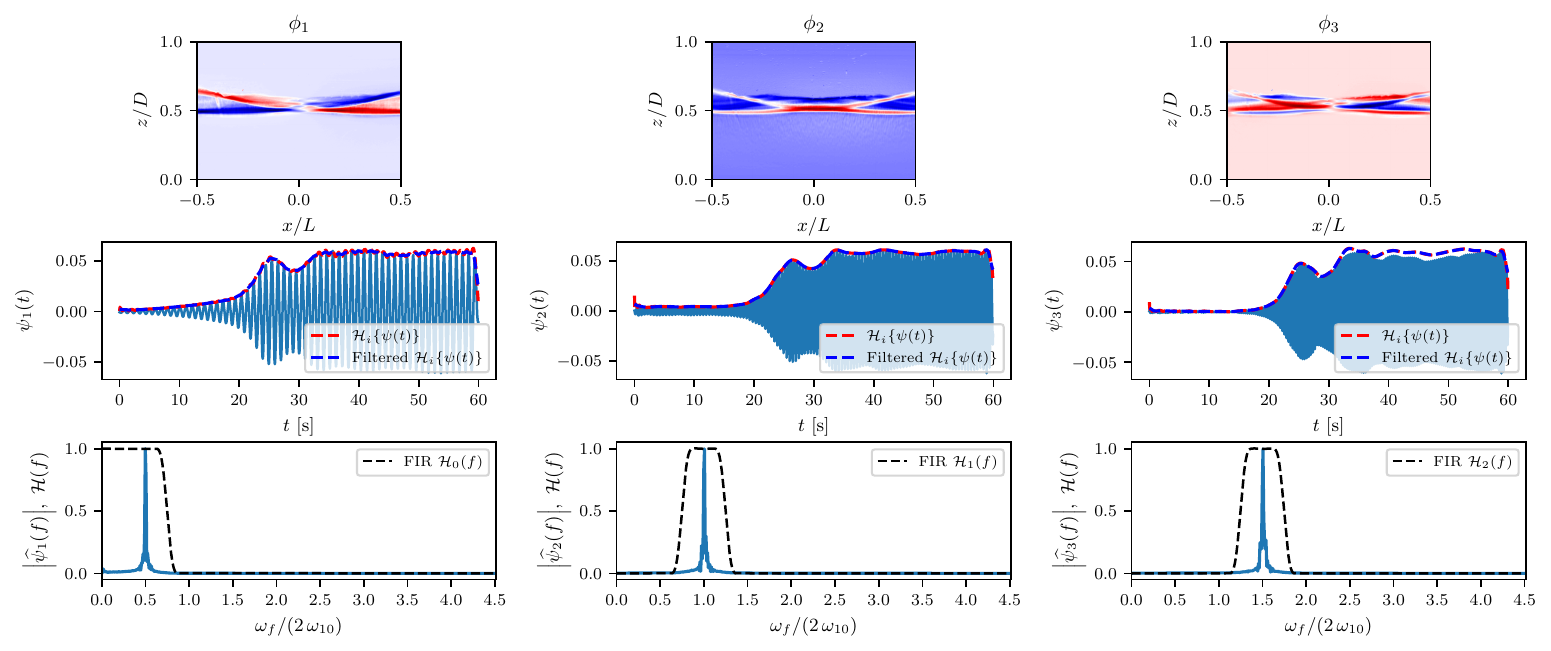}
    \label{fig:p2_m1_med_prototype}
}
\subfigure[Prototype $p_3$: $A_f/R = 0.33, \,\, \omega_f/(2\,\omega_{1,0}) = 1.00$.]{
    \includegraphics[width=1\textwidth]{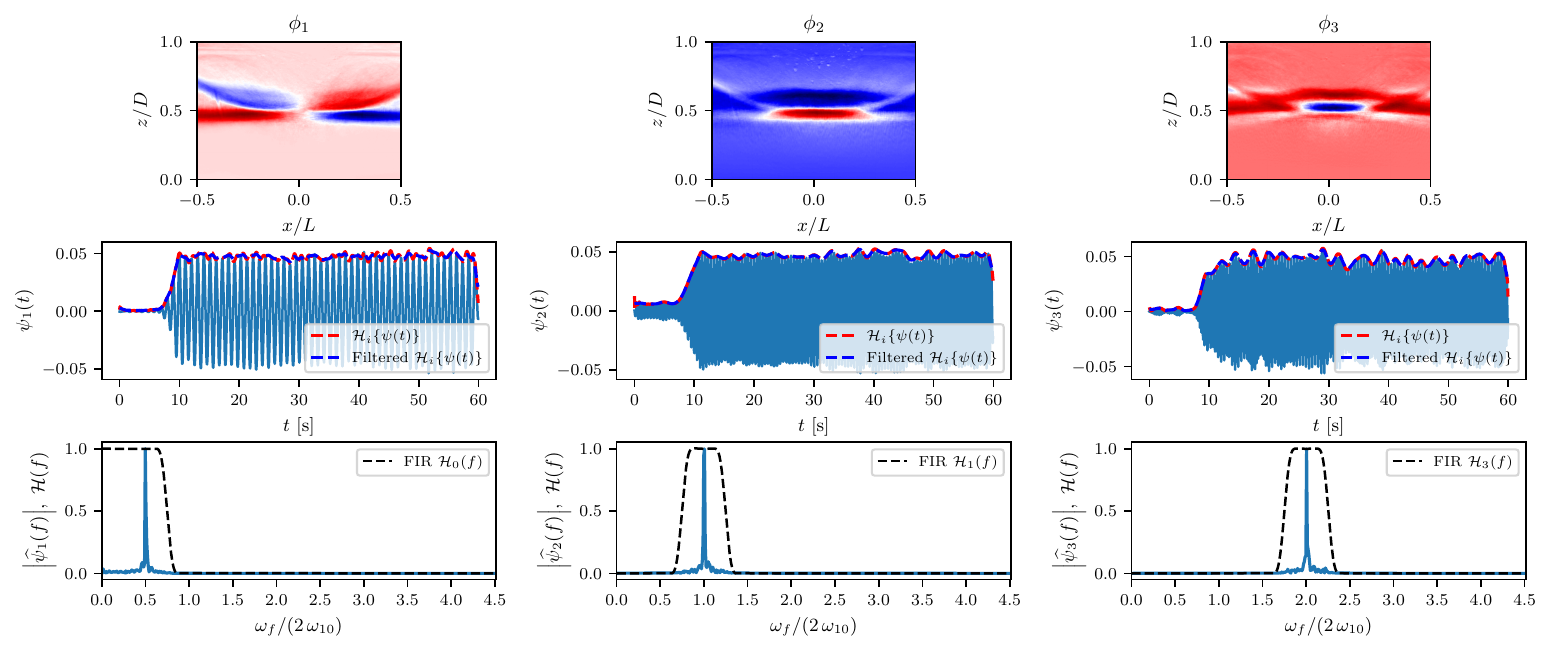}
    \label{fig:p3_m1_wb_prototype}
}
\caption{mPOD spatial structures $\boldsymbol{\phi}_{m_k}$, normalized frequency spectra $|\hat{\boldsymbol{\psi}}_{m_k}|$ with overlaid FIR filters $\mathcal{H}_m$, and temporal structures $\boldsymbol{\psi}_{m_k}$ for three prototype references $p_1$ to $p_3$, each characterized by longitudinal sloshing dynamics $(m = 1, n = 0)$.}
\label{fig:longitudinal_prototypes}
\end{figure*}

\begin{figure*}[htp!]
\centering
\subfigure[Prototype $p_4$: $A_f/R = 0.40, \,\, \omega_f/(2\,\omega_{1,0}) = 1.17$.]{
    \includegraphics[width=1\textwidth]{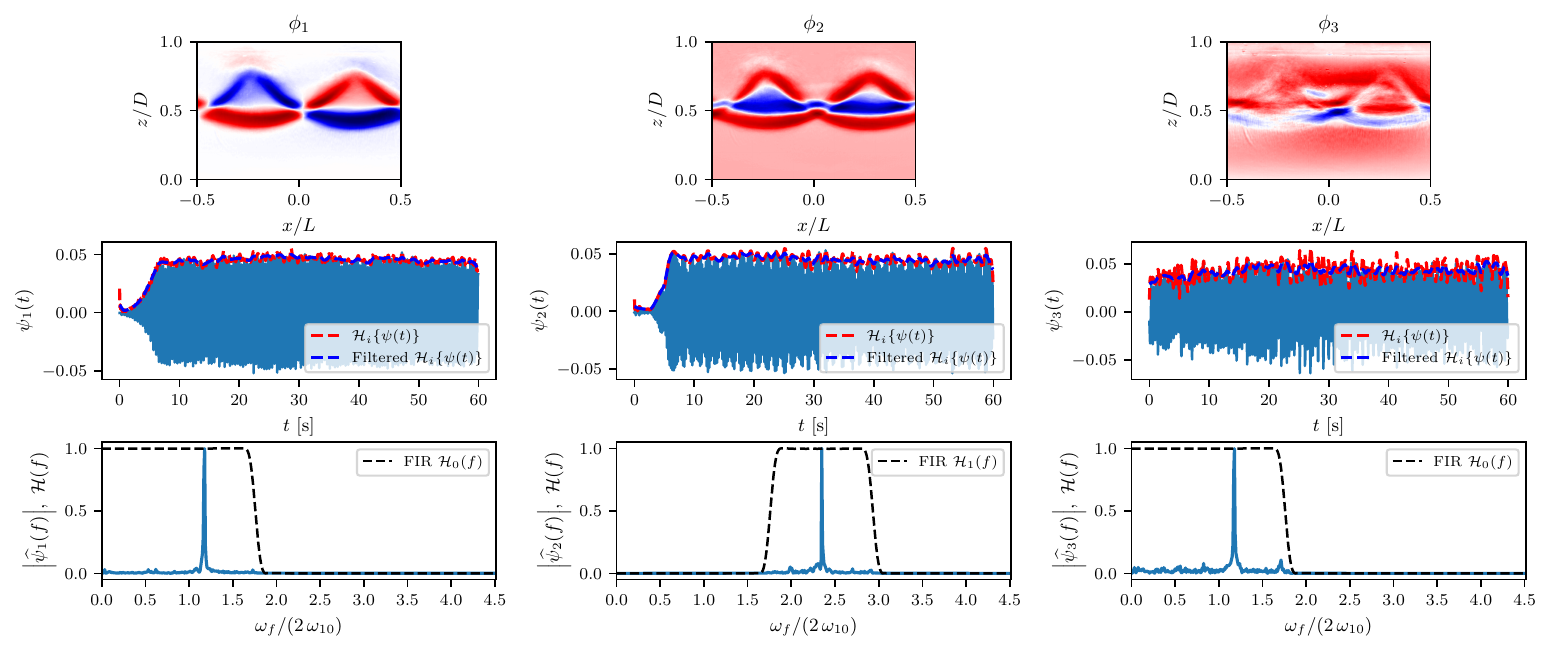}
    \label{fig:p4_m3_prototype}
}
\subfigure[Prototype $p_5$: $A_f/R = 0.41, \,\, \omega_f/(2\,\omega_{1,0}) = 0.91$.]{
    \includegraphics[width=1\textwidth]{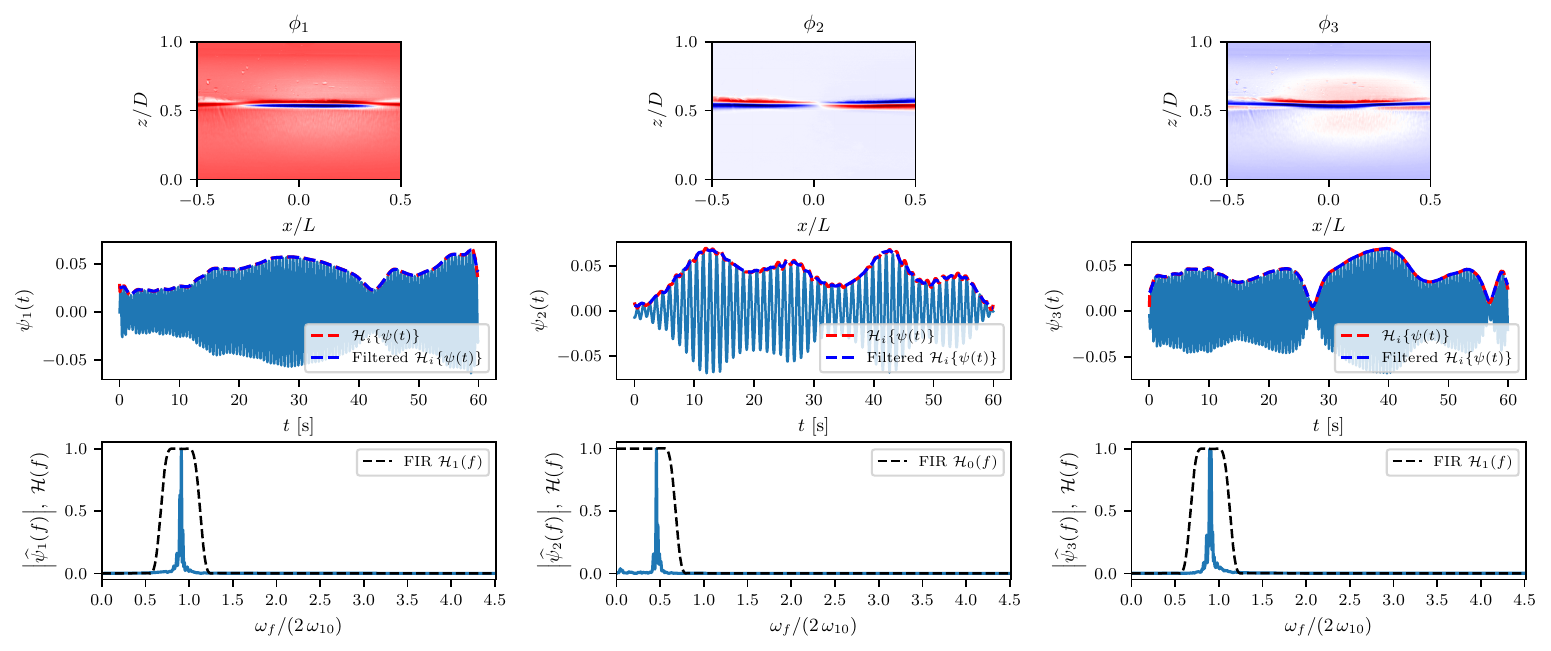}
    \label{fig:p5_mix_low_prototype}
}
\subfigure[Prototype $p_6$: $A_f/R = 0.25, \,\, \omega_f/(2\,\omega_{1,0}) = 1.17$.]{
    \includegraphics[width=1\textwidth]{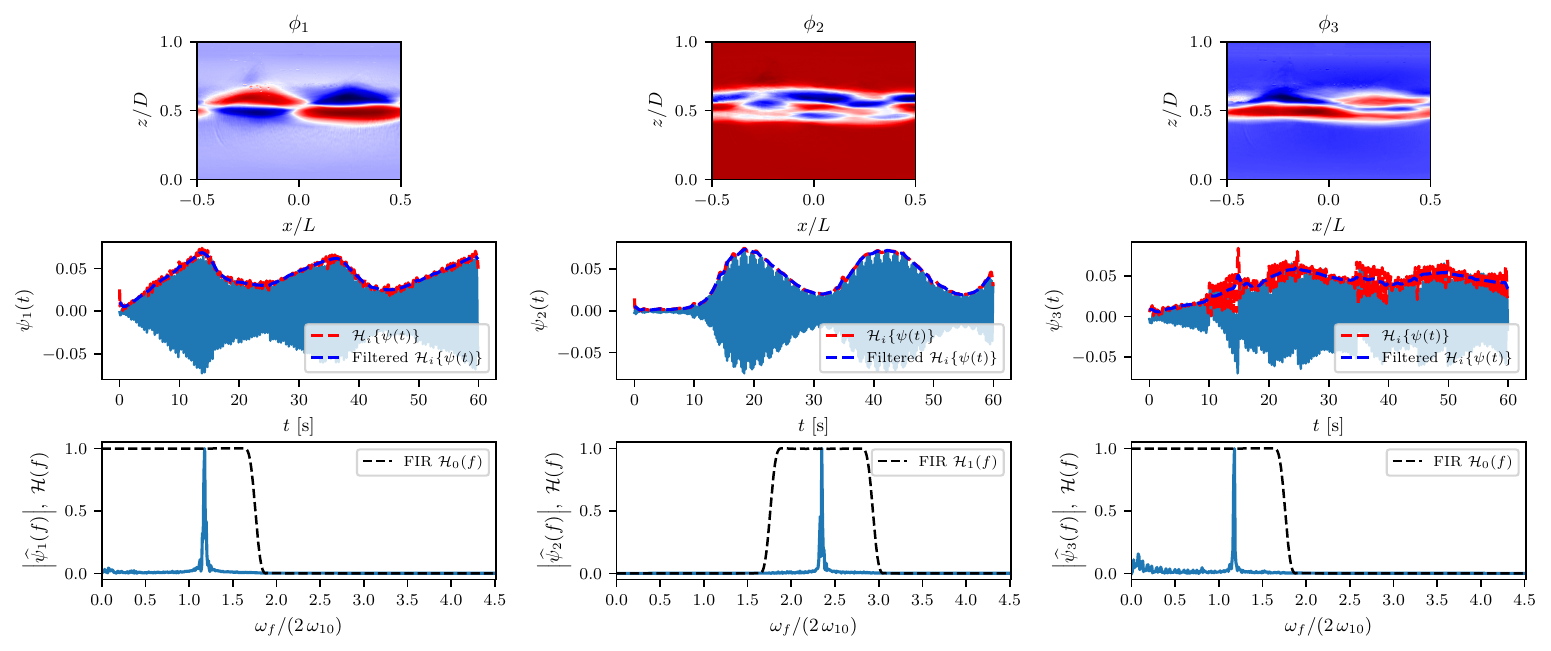}
    \label{fig:p6_mix_high_prototype}
}
\caption{Same as Fig. \ref{fig:longitudinal_prototypes} but considering the prototypes $p_4$ to $p_6$, characterized by high-order longitudinal dynamics $(m = 2, n = 0)$ and $(m = 3, n = 0)$, with mixed-mode conditions in $p_5$ and $p_6$.}
\label{fig:mixing_prototypes}
\end{figure*}

\begin{figure*}[t!]
\centering
\subfigure{
    \includegraphics[width=0.96\textwidth]{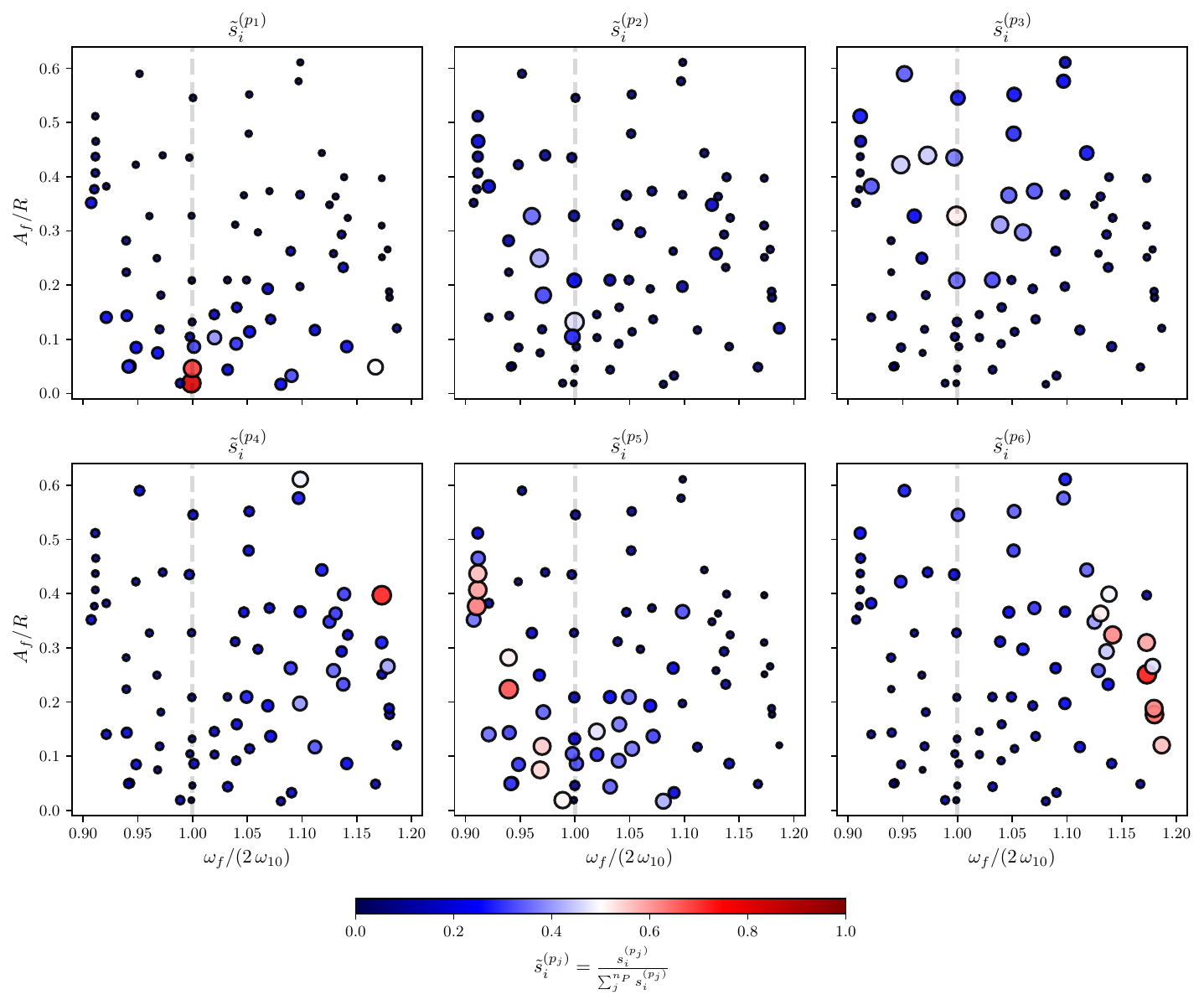}
    \label{fig:similarity_score}
}
\caption{Prototypes (see Figs. \ref{fig:longitudinal_prototypes} and \ref{fig:mixing_prototypes}) similarity maps in the dimensionless forcing space defined by the frequency ratio \( \omega_f / (2\,\omega_{1,0}) \) and amplitude ratio \( A_f / R \). Each marker represents a test condition, with color indicating the normalized similarity score \( \tilde{s}_i^{(p_j)} = s_i^{(p_j)} / \sum_{j=1}^{n_P} s_i^{(p_j)} \) defined in \eqref{eq:prototype_presence} and size proportional to the score.}
\label{fig:similarity_score_map}
\end{figure*}

The image sequences were analyzed over $t \in [0,60]\,\mathrm{s}$ at a down-sampled rate of $17.5\,\mathrm{Hz}$ ($n_t=1050$), after spatial cropping to center the tank ($n_x=1900$, $n_z=1185$). A set of six representative prototypes ($n_P=6$), each described by its three leading mPOD modes ($m_j=3$), was selected to span the range of observed sloshing behaviors near $2\omega_{1,0}$. These reference cases define the basis $\Phi_{ref}$ onto which all test realizations are projected to extract energy-based features for subsequent clustering (Sec.~\ref{sec:clustering_results}).  

Figs.~\ref{fig:longitudinal_prototypes} and~\ref{fig:mixing_prototypes} present the selected prototypes, each displayed with its spatial mode shapes $\boldsymbol{\phi}_{m_k}^{(p_j)}$, temporal coefficients $\boldsymbol{\psi}_{m_k}^{(p_j)}(t)$ and their Hilbert envelopes, as well as the corresponding frequency spectra $|\hat{\boldsymbol{\psi}}_{m_k}^{(p_j)}|$ and the transfer functions $\mathcal{H}_m$ of the filters that identify each mode. The forcing amplitude and frequency for every test are indicated in the subplot labels. Together, these examples illustrate regimes ranging from a nearly flat free surface at low forcing amplitudes to pronounced longitudinal $(m=1,\,n=0)$ dynamics and higher-order (e.g., $m=3$) or mixed-mode responses, forming the foundation for the reduced-order classification.

All prototypes in Fig.~\ref{fig:longitudinal_prototypes} correspond to the resonance condition $\omega_f = 2\omega_{1,0}$. Prototype~$p_1$, shown in Fig.~\ref{fig:p1_stable_prototype} and having $A_f/R=0.02$, represents a stable free-surface response. The spatial structure $\boldsymbol{\phi}_1^{(p_1)}$ exhibits only faint vestiges of the longitudinal mode, while the temporal coefficient $\boldsymbol{\psi}_1^{(p_1)}(t)$ decays exponentially in time. The corresponding spectrum reveals a weak component at the longitudinal-mode frequency, confirming that the system remains below the instability threshold and that the interface returns to rest after the excitation. Prototype~$p_2$, with $A_f/R = 0.13$, also exhibits a strong longitudinal response, but in this case the oscillation grows toward a well-established periodic motion. The spatial structure $\boldsymbol{\phi}_1^{(p_2)}$ shows a single nodal line at the tank mid-plane and a nearly linear interface profile, characteristic of the first antisymmetric longitudinal mode. The temporal coefficient $\boldsymbol{\psi}_1^{(p_2)}$ reaches a steady amplitude, indicating sustained subharmonic sloshing consistent with the onset of the $(m=1,\,n=0)$ resonance. Prototype~$p_3$, with $A_f/R = 0.33$, with qualitatively similar longitudinal dynamics as $p_2$ but with larger amplitude. The system converges toward a fully developed limit cycle much faster, as evidenced by the rapid growth of the leading temporal mode $\boldsymbol{\psi}_1^{(p_3)}$. The corresponding spatial structure $\boldsymbol{\phi}_1^{(p_3)}$ retains a single nodal line at mid-span but exhibits a noticeably curved interface, reflecting stronger nonlinear effects associated with the higher forcing amplitude.

Prototype~$p_4$, with $A_f/R = 0.42$ and $\omega_f/(2\omega_{1,0}) = 1.17$, shown in Fig.~\ref{fig:p4_m3_prototype}, marks the onset of more complex standing-wave behavior. The spatial structure $\boldsymbol{\phi}_1^{(p_4)}$ exhibits additional nodes near the tank sidewalls and antinodes at $x/L \approx 0.25$ and $x/L \approx 0.75$, resembling a higher-order $(m=3,n=0)$-type pattern. The temporal coefficient $\boldsymbol{\psi}_1^{(p_4)}(t)$ displays amplitude modulation and intermittent growth–decay cycles, indicating mode interaction and partial energy transfer between the fundamental longitudinal oscillation and its higher harmonics. Prototype~$p_5$, with $A_f/R = 0.52$ and $\omega_f/(2\omega_{1,0}) = 1.04$, shown in Fig.~\ref{fig:p5_mix_low_prototype}, represents a strongly nonlinear regime characterized by pronounced wave steepening and intermittent breaking. The spatial mode $\boldsymbol{\phi}_1^{(p_5)}$ displays a distorted longitudinal pattern with asymmetric crest–trough geometry but overall smaller amplitude than in $p_4$, suggesting partial energy redistribution among competing modes. The temporal coefficient $\boldsymbol{\psi}_1^{(p_5)}$ exhibits irregular amplitude fluctuations and intermittent bursts, while the corresponding spectrum is significantly broadened, reflecting the coexistence of multiple harmonics and enhanced mode coupling.

Finally, prototype~$p_6$, with $A_f/R = 0.62$ and $\omega_f/(2\omega_{1,0}) = 0.94$, shown in Fig.~\ref{fig:p6_mix_high_prototype}, represents a mixed-mode regime with strong nonlinear modulation. The spatial mode $\boldsymbol{\phi}_1^{(p_6)}$ retains a weakened longitudinal pattern with intermittent distortions, while the temporal coefficient $\boldsymbol{\psi}_1^{(p_6)}$ shows pronounced amplitude modulation. The spectrum remains harmonic, with discrete peaks and sidebands indicating coherent oscillations modulated by nonlinear interactions.

Overall, the six prototypes span the main sloshing behaviors --- from stable waves ($p_1$) to subharmonic longitudinal responses ($p_2$–$p_3$) and mixed higher-order modes away from resonance. Assuming this reference basis is sufficiently rich, each test realization can be projected onto it (see~\eqref{eq:projection_A_proj}) to compute the activation energies used for regime clustering (see~\eqref{eq:feature_vector}).


\subsection{Similarity maps} \label{sec:similarity_bubble_maps}

Figure~\ref{fig:similarity_score_map} presents the similarity maps across the dimensionless forcing space for the six prototypes described in the previous subsection. The similarity score at each point is computed in step~3 of the proposed workflow (see Fig.~\ref{fig:flowchart_methodology}) from the normalized energy signature obtained through the filtered autoencoder formulation. The projection in~\eqref{eq:projection_A_proj} was performed on a low-order reconstruction of each video, retaining the first $n_R = 30$ mPOD modes --- thereby using the mPOD basis as both a decomposition and filtering tool.

Distinct structural patterns emerge, with clear regions of influence corresponding to the activation of each prototype. This confirms the effectiveness of the proposed similarity metric and highlights the importance of a representative prototype selection. Prototype~$p_1$ exhibits the highest similarity scores in the lower region of the map, delineating the boundary of stable free-surface conditions. As the forcing amplitude increases, prototype~$p_2$ gradually dominates, corresponding to the onset of longitudinal oscillations slightly below the parametric resonance frequency. A sharp transition then occurs toward the high-amplitude region represented by prototype~$p_3$, as reflected in the elevated scores $\tilde{s}_i^{(p_3)}$. 

Prototype~$p_5$, characterized by mixed low-amplitude longitudinal dynamics, shows partial overlap with prototypes~$p_1$ and~$p_2$, suggesting shared behavior. This overlap implies that a reduced reference set of four or five prototypes may suffice to capture the low-amplitude macro-sloshing regime. Nevertheless, the smooth progression between stable and weakly oscillatory states underscores the importance of a refined prototype library for resolving subtle dynamic transitions, particularly near $A_f/R \approx 0.1$.

In contrast, prototypes~$p_3$, $p_4$, and~$p_6$ remain well separated, each defining a distinct region of influence. The similarity distributions $\tilde{s}_i^{(p_4)}$ and $\tilde{s}_i^{(p_6)}$, both associated with higher-order longitudinal dynamics ($m=3$), indicate the emergence of a secondary instability region characterized by mode interaction and mixed-mode responses. Interestingly, this region develops along a slope nearly identical to that of $\tilde{s}_i^{(p_2)}$ and $\tilde{s}_i^{(p_5)}$, suggesting the formation of a second instability tongue.

\begin{figure*}[ht!]
\centering
\subfigure{
    \includegraphics[width=0.5\textwidth]{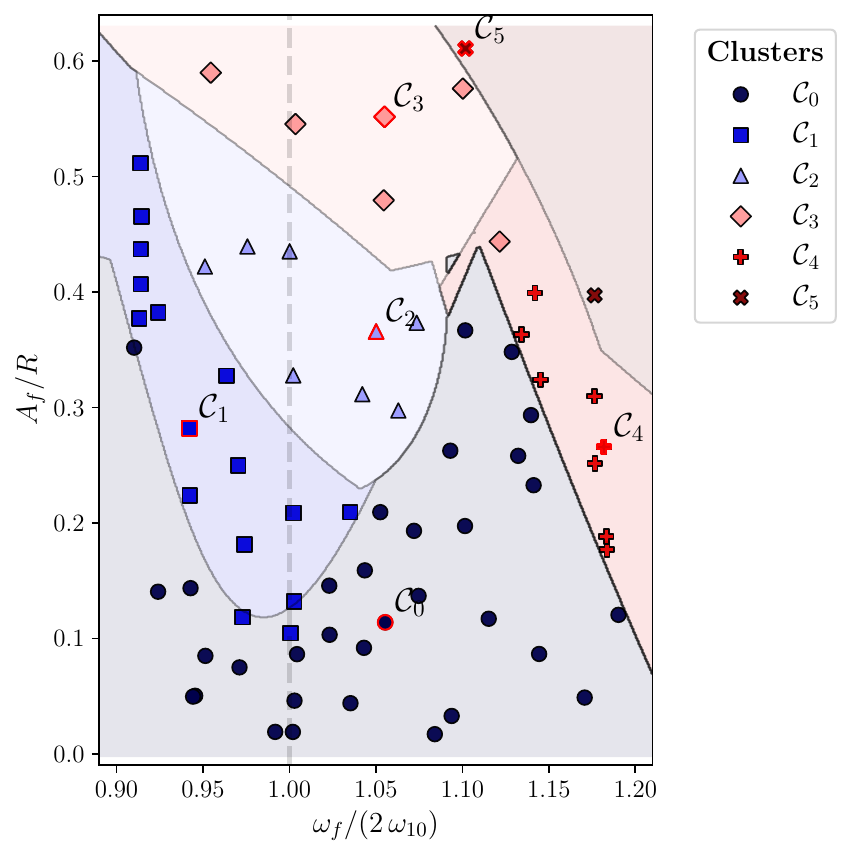}
    \label{fig:k_means_map}
}
\subfigure{
    \includegraphics[width=0.96\textwidth]{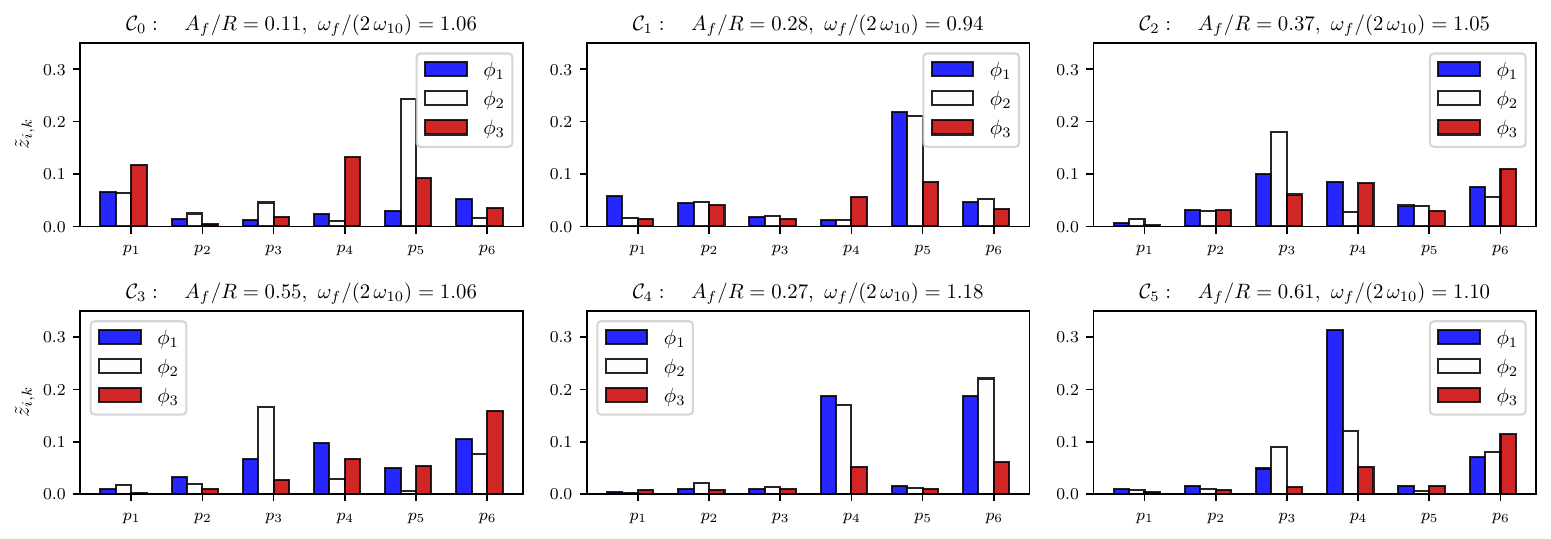}
    \label{fig:l2_norms_centroid}
}

\caption{Classification results for $H_l/D = 0.5$, with the data subdivided into six distinct clusters: \( \{ \mathcal{C}_0, \ldots, \mathcal{C}_{5} \} \), equaling the number of reference prototypes $p_j = p_6$ (see Sec. \ref{sec:prototype_library}). The experimental space was explored in $A_f / R \in [0.02, 0.62]$, with varying acceleration $A_f\omega_f^2/g \in  \pm [0.02, 0.77]$, while the dimensionless forcing frequency was investigated over the range of $\omega_f /(2 \, \omega_{1,0}) \in [0.90, 1.20]$. The normalized energy signature of each feature vector component, $\tilde{z}_{i,k}$, is shown for every cluster medoid.}
\label{fig:k_means_cluster}
\end{figure*}
\subsection{Clustering and regime classification} \label{sec:clustering_results}

This subsection outlines the results obtained from the prototype clustering and classification procedure. In Fig.~\ref{fig:k_means_cluster}, the test points in the dimensionless forcing diagram, represented as $\left( A_f/R; \,\omega_f / (2 \, \omega_{1,0})\right)$, are clustered according to the output from the unsupervised $k$-means algorithm: \( \{ \mathcal{C}_0, \dots, \mathcal{C}_{5} \} \). The number of clusters is treated as a hyperparameter and is set equal to the number of prototypes, resulting in six distinct clusters. To support the analysis, bar plots are shown for each cluster medoid, illustrating the relative contribution of each component in the feature vector, \mbox{$\tilde{z}_{i,k} = z_{i,k} / \sum_{j=1}^{n_b} z_{i,j}, \quad k = 1, \dots, n_b,$} with entries grouped by prototype.

Cluster $\mathcal{C}_1$, centered at $A_f/R = 0.28$, $\omega_f / (2 \, \omega_{1,0}) = 0.94$, captures points in the lower region of the parametric resonance condition along with test conditions at moderately high forcing amplitudes $0.15 \leq A_f / R \leq 0.55$, yet mostly constrained to sub-resonant frequencies with $\omega_f / (2 \, \omega_{1,0}) \leq 1.00$. The bar plot with the relative similarity $\tilde{{z}}_{i,k}$ identifies $p_5$ — a mixed low-amplitude longitudinal case — as the dominant prototype (see Fig.~\ref{fig:p5_mix_low_prototype}) in this region.

Cluster $\mathcal{C}_3$, with medoid at ($A_f/R = 0.55$, $\omega_f / (2 \, \omega_{1,0}) = 1.06$), identifies high-amplitude forcing ($A_f/R \geq 0.45$) near resonance. Here, $\boldsymbol{\phi}_2$ from $p_3$ shows a strong response, with a secondary contribution from $p_6$. While $p_3$ reflects wave-breaking in the $m=1$ mode (Fig.~\ref{fig:p3_m1_wb_prototype}), the presence of $p_6$ suggests that at $\omega_f / (2 \, \omega_{1,0}) \geq 1.05$, higher-order longitudinal modes like $m=3$ may emerge, though potentially under-sampled. Notably, $\mathcal{C}_2$ closely resembles $\mathcal{C}_3$, with a key difference: feature-wise, $p_3$ has a larger impact than $p_6$. Therefore, the region defined by $0.32 \leq A_f / R \leq 0.45$ and $0.95 \leq \omega_f / (2 \, \omega_{1,0}) \leq 1.07$ can be categorized as conditions of pure wave-breaking longitudinal waves. In contrast, at higher forcing frequencies --- particularly near the medoid of $\mathcal{C}_3$ --- the effects of higher-order longitudinal modes may become significant.

Cluster $\mathcal{C}_0$ shows a mixed signature, primarily influenced by $p_5$ and secondarily by $p_4$ and $p_1$. Notably, $\boldsymbol{\phi}_3$ from $p_4$ contributes over 15\%, while $\boldsymbol{\phi}_2$ and $\boldsymbol{\phi}_3$ from $p_5$ contribute over 30\% of the total projected energy. These higher-order modes from prototype $p_5$ resemble those of $p_1$ but exhibit greater amplitudes. Clearly, with over 30 samples identified, one can distinguish a strong boundary between stable free-surface dynamics (see Fig.~\ref{fig:p1_stable_prototype}) and sloshing conditions (i.e., $\mathcal{C}_0\ \text{vs.}\ \mathcal{C}_{\neg 0}$) located in the lower part of the diagram. 

The histogram reveals that cluster $\mathcal{C}_4$ emerges from the interplay between prototypes $p_4$ and $p_6$. While $p_4$ represents pure longitudinal motion with $m=3$, $p_6$ exhibits a mixed response involving both longitudinal ($m=3$) and transverse ($n=1$) modes. This supports the existence of a distinct instability region at $\omega_f / (2 \, \omega_{1,0}) \geq 1.20$. A detailed characterization of this regime, however, lies beyond the scope of this study.

Finally, for cluster~$\mathcal{C}_5$, limited sampling in this high-amplitude and high-frequency region, together with the mechanical constraints of the sloshing table, resulted in only two data points. One of these corresponds to prototype~$p_4$ at \(A_f/R = 0.40\) and \(\omega_f/(2\,\omega_{1,0}) = 1.17\), situated near the lower bound of this cluster. 

Notably, within the iso-frequency range \(1.15 \leq \omega_f / (2\,\omega_{1,0}) \leq 1.20\), higher-order longitudinal motion becomes dominant, giving rise to a secondary instability region associated with the \(m=3\) mode. This region originates from nonlinear coupling between longitudinal, transverse, and oblique components. This interaction produces mixed modal responses, a phenomenon characteristic of horizontally oriented cylindrical tanks where near-degenerate natural frequencies promote mode competition (see Fig.~\ref{fig:natfreq_vs_fill}). The secondary harmonic branch associated with these modes develops near the primary subharmonic instability band of the longitudinal \(m=1\) mode, resulting in overlapping resonance tongues. As a result, the excitation frequency governs energy exchange between interacting modes, determining whether mixed or pure longitudinal responses prevail. This coupled-resonance behavior is consistent with the nonlinear Mathieu framework of~\citet{el-dib_nonlinear_2001}, which predicts the simultaneous occurrence of parametric instabilities and nested resonance regions. It also aligns with experimental observations of mode competition and mixed states in parametrically forced surface waves reported by~\citet{GOLLUB_surface_1989} and more recently by~\citet{colville_faraday_2025}.

\subsection{Parametric sloshing maps and fill level dependence} \label{sec:sloshing_maps}

\begin{figure*}[ht!]
\centering
\includegraphics[width=0.92\textwidth]{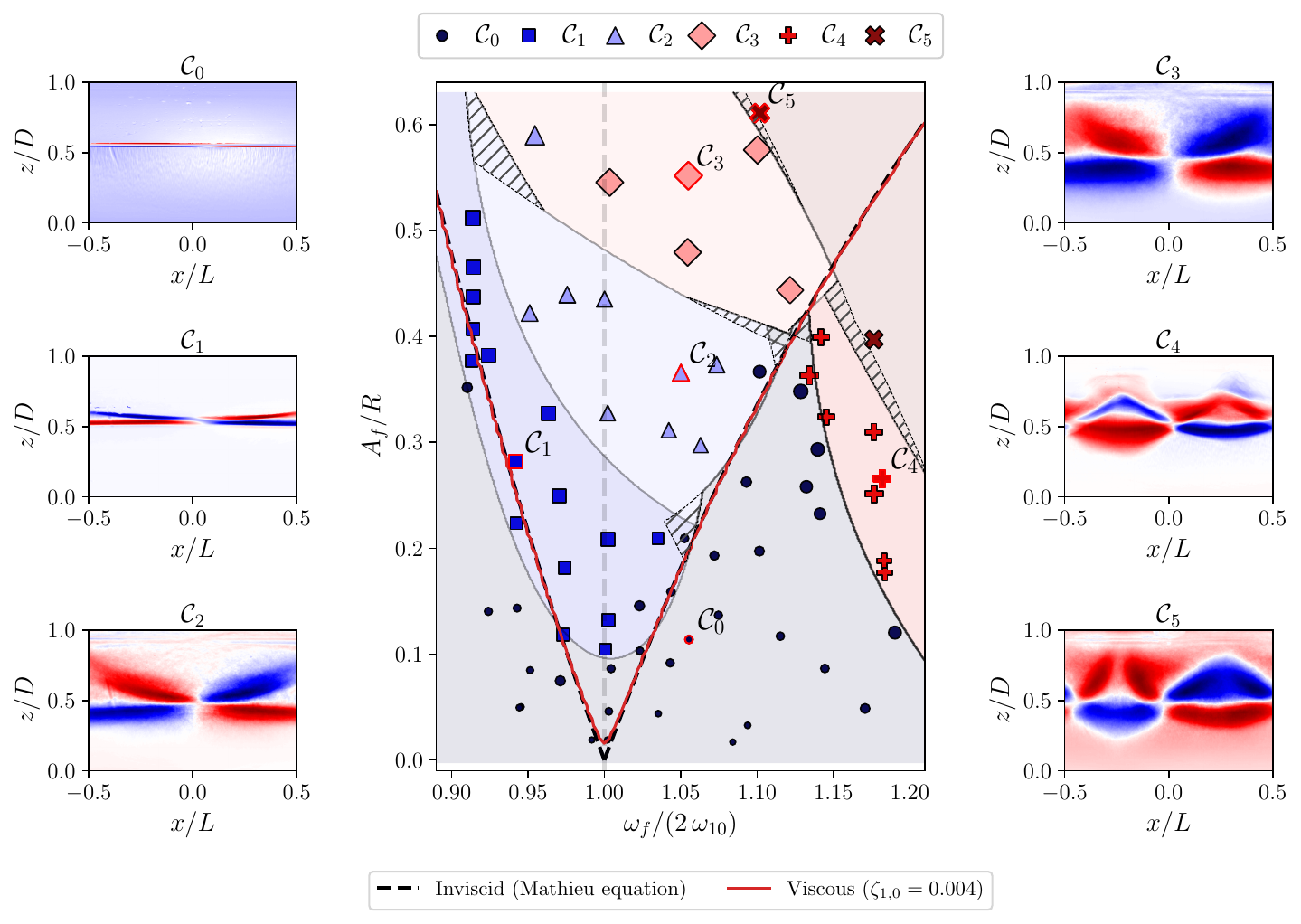}

\caption{Phase diagram of the dimensionless forcing amplitude, $A_f/R$, versus the frequency ratio, $\omega_f /(2 \, \omega_{1,0})$, showing the stability boundaries in the vicinity of twice the fundamental sloshing eigenfrequency $\omega_{1,0}$ for a fill ratio of $H_l/D = 0.50$. Marker size highlights the sloshing-induced mixing intensity, $I_i$, quantified through the overall modal energy of the first thirty mPOD modes. For each cluster, the medoid’s highest-energy spatial structure, $\boldsymbol{\phi}_1$, is displayed: $\mathcal{C}_0$ – stable free-surface; $\mathcal{C}_1$ – low-amplitude longitudinal motion; $\mathcal{C}_2$ – longitudinal mode $m=1$; $\mathcal{C}_3$ – wave-breaking longitudinal mode $m=1$; $\mathcal{C}_4$ – mixed high-order dynamics; $\mathcal{C}_5$ – longitudinal mode $m=3$. Parametric-resonance predictions based on the Mathieu equation, in both the undamped ($\zeta_{1,0}=0$) and damped ($\zeta_{1,0}=0.004$) cases, are represented by the neutral-stability boundaries.} 

\label{fig:final_regime_map_50}
\end{figure*} 

\begin{figure*}[ht!]
\centering
\includegraphics[width=0.9\textwidth]{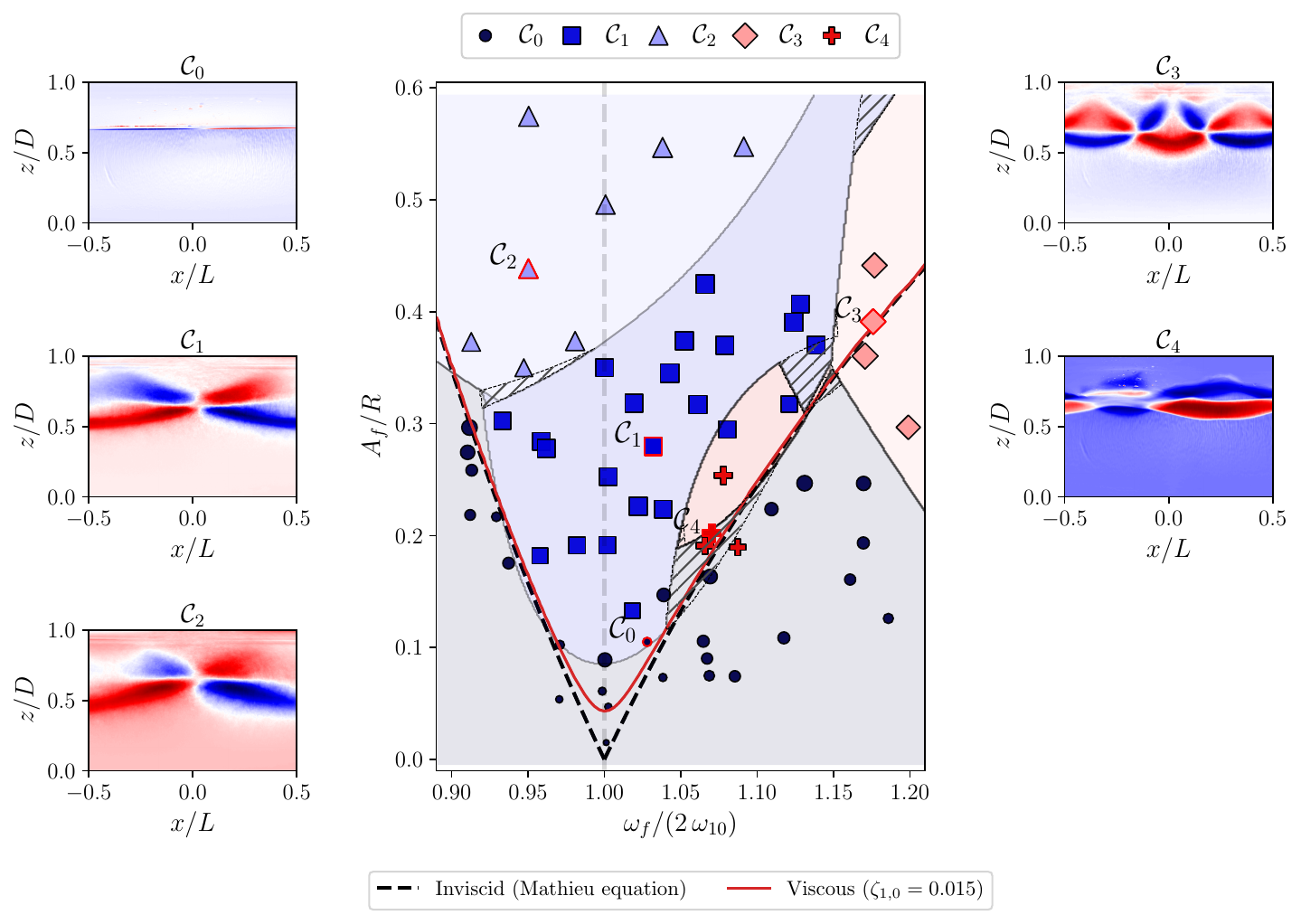}

\caption{Same as Fig.~\ref{fig:final_regime_map_50}, but for a fill ratio of $H_l/D = 0.67$. The clustered data separate into: $\mathcal{C}_0$ – stable free surface; $\mathcal{C}_1$ – pure longitudinal $m=1$; $\mathcal{C}_2$ – breaking $m=1$; $\mathcal{C}_3$ – pure higher-order $m=4$; $\mathcal{C}_4$ – mixed high-order dynamics. Neutral-stability boundaries from the inviscid and viscous ($\zeta_{1,0}=0.015$) Mathieu solutions indicate the parametric-resonance region.} 

\label{fig:final_regime_map_67}
\end{figure*} 

This subsection presents the final sloshing-regime maps for the tested fill ratios (see Table~\ref{tab:experimental_matrix}). These maps are obtained through the iterative labeling strategy described in Section~\ref{sec:labeling_data}, in which the reference basis (see~\eqref{eq:reference_basis}) is progressively updated using cluster medoids until the convergence criteria of Subsec.~\ref{sec:labeling_data} are met. This process improves cluster separation and reduces sensitivity to the initial hand-picked prototype set.

The sloshing intensity $I_i$ at each operating point is represented through the marker size, computed from the total modal energy of the first $n_R = 30$ mPOD modes as
\[
I_i = \sqrt{\sum_{r=1}^{30} \sigma_r^{2}}.
\]
Marker sizes follow the same linear mapping used in Subsec.~\ref{sec:similarity_bubble_maps}. This metric provides a direct measure of the fluctuation energy contained in the dominant sloshing structures, allowing regions of strong activity to be visually emphasized within the parameter space.

Figure~\ref{fig:final_regime_map_50} shows the refined regime map for $H_l/D = 0.50$, building upon the initial clusters in Fig.~\ref{fig:k_means_map}. Uncertain near-boundary regions are indicated using the hatched rendering described in Subsec.~\ref{sec:SVM}. Convergence was achieved after three iterations, with an ARI history $\{0.97,\,0.99\}$~\citep{ARI_article}. The final medoids remained consistent with those from the first iteration, although convergence was triggered by the ARI threshold rather than the prototype-overlap condition. The low intra-cluster variance confirms the robustness of the final clusters and the distinct nature of the six regimes defined by the prototype library (Figs.~\ref{fig:longitudinal_prototypes}–\ref{fig:mixing_prototypes}).

\begin{figure}[b!]
\centering
\includegraphics[width=0.50\columnwidth]{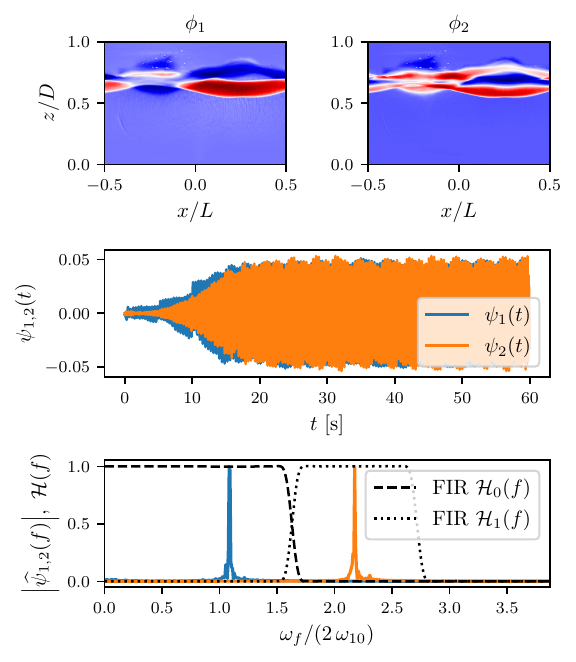}
\caption{mPOD modes $\boldsymbol{\phi}_{1,2}$, temporal coefficients $\boldsymbol{\psi}_{1,2}$, and spectra $|\hat{\boldsymbol{\psi}}_{1,2}|$ for $\mathcal{C}_4$ at $H_l/D=0.67$ (see Fig. \ref{fig:final_regime_map_67}).} 
\label{fig:67_psi_phi}
\end{figure} 

\begin{figure*}[ht!]
\centering
\includegraphics[width=0.9\textwidth]{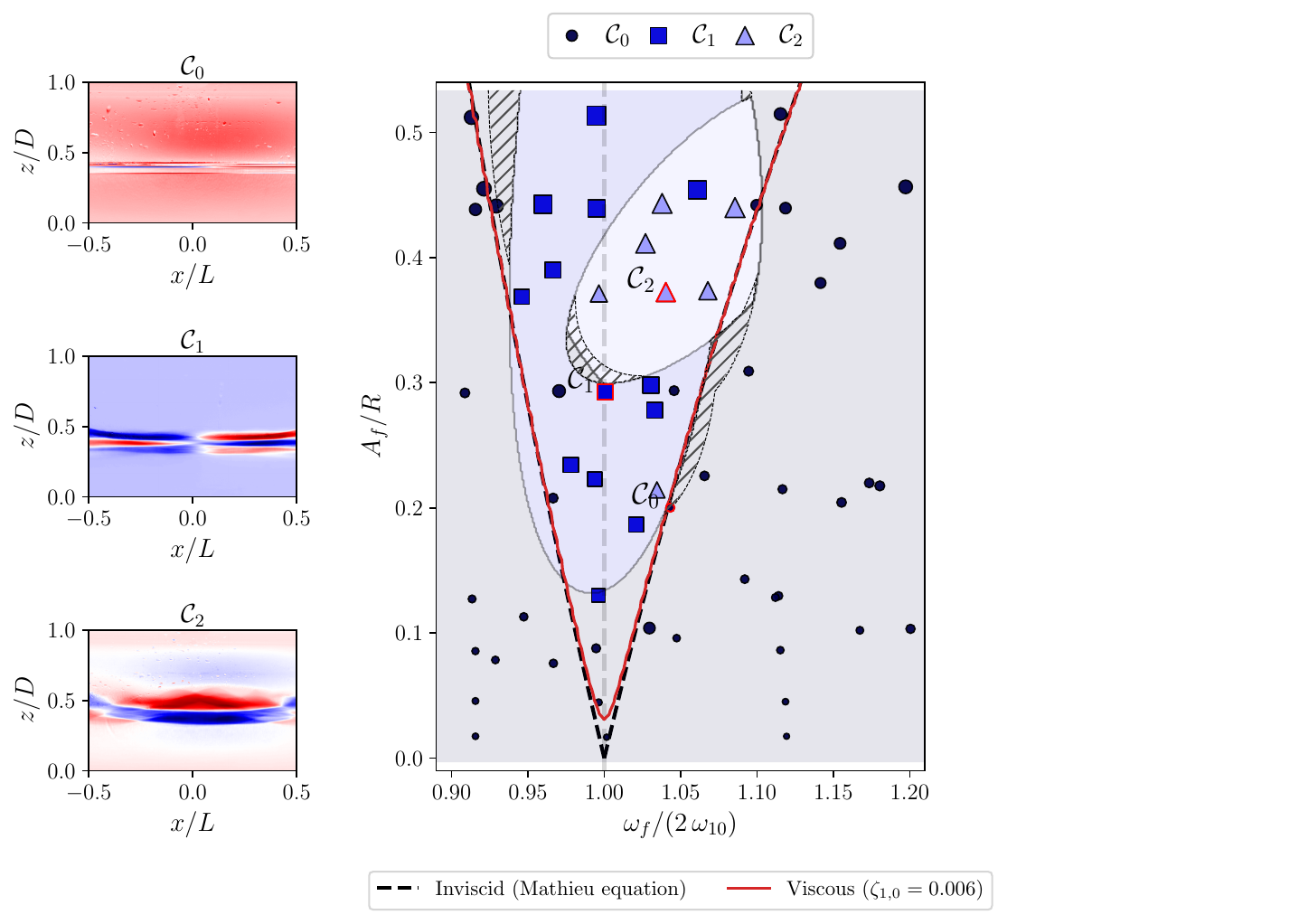}
\caption{Same as Fig.~\ref{fig:final_regime_map_50}, but for $H_l/D = 0.40 \pm 0.04$, with the data subdivided into three clusters: $\mathcal{C}_0$ — stable free-surface dynamics; $\mathcal{C}_1$ — longitudinal mode $m=1$; and $\mathcal{C}_2$ — moderate-amplitude longitudinal motion. Parametric resonance, computed from the inviscid and viscous Mathieu equations, is shown through the corresponding neutral-stability boundaries.}
\label{fig:final_regime_map_40}
\end{figure*}

\begin{figure}[b!]
\centering
\includegraphics[width=0.50\columnwidth]{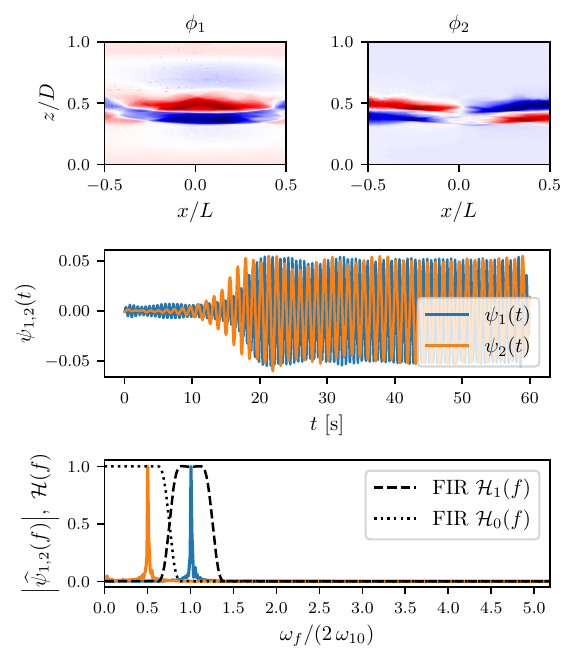}
\caption{mPOD modes $\boldsymbol{\phi}_{1,2}$, temporal coefficients $\boldsymbol{\psi}_{1,2}$, and spectra $|\hat{\boldsymbol{\psi}}_{1,2}|$ for $\mathcal{C}_2$ at $H_l/D=0.40$ (see Fig. \ref{fig:final_regime_map_40}).} 
\label{fig:40_psi_phi}
\end{figure} 

The dominant spatial structure $\boldsymbol{\phi}_1$ of each medoid is shown in the surrounding panels of Fig.~\ref{fig:final_regime_map_50}. Cluster $\mathcal{C}_2$ occupies the central region of the map and corresponds to the pure longitudinal $m=1$ mode. Its butterfly-like curvature, with slight doming near $x/L=\pm0.5$, highlights the potential for interface breakup and liquid ejection near the spherical tank heads. A similar pattern appears in clusters $\mathcal{C}_3$ and $\mathcal{C}_1$, reflecting a systematic increase in modal amplitude with forcing. At small forcing amplitudes the system displays finite subharmonic responses characteristic of the weakly damped linear regime. As the forcing increases, the oscillations saturate due to nonlinear effects, and at the highest amplitudes ($A_f/R \gtrsim 0.45$) the motion becomes constrained by the tank geometry as the interface impinges on the bulkheads and cylindrical walls. In this high-forcing regime, vigorous mixing is induced through wave run-up, liquid ejection, and repeated wave breaking.

Clusters $\mathcal{C}_4$ and $\mathcal{C}_5$, appearing for $\omega_f/(2\omega_{1,0}) \geq 1.12$, exhibit higher-order longitudinal responses with $m=3$, consistent with the prototype dynamics shown in Figs.~\ref{fig:p4_m3_prototype} and~\ref{fig:p6_mix_high_prototype}. These regimes involve geometry-limited wave amplitudes and mixed-mode interactions, as discussed in Subsec.~\ref{sec:clustering_results}.

The maps in Fig.~\ref{fig:final_regime_map_50} also include the neutral stability boundaries predicted by the inviscid ($\zeta_{1,0}=0$) and viscous ($\zeta_{1,0}=0.004$) Mathieu-equation solutions of Sec.~\ref{sec:Mathieu_eq}. These boundaries are centered at the experimental value $2\omega_{1,0}$, partitioning stable (free-surface) and unstable (sloshing) regions. Classical results~\citep{benjamin_ursell_1954, brand_parametrically_1965} show that viscous dissipation sets a critical excitation amplitude below which the free surface remains stable, in agreement with the present dataset. A marked increase in wave amplitude, captured here by cluster $\mathcal{C}_1$, occurs only for $A_f/R \gtrsim 0.1$ ($A_f \gtrsim \SI{6.7}{mm}$). Overall, the predicted instability envelope aligns closely with the experimental classification, with most $\mathcal{C}_0$ (stable) points lying outside the Mathieu region. The physics-informed framework therefore reinforces the consistency of the regime map, even though the Mathieu model —-- with only linear damping --- slightly underpredicts the excitation amplitude required for instability.

Figure~\ref{fig:final_regime_map_67} presents the iteratively refined regime map for the fill ratio $H_l/D = 0.67 \pm 0.04$. Convergence was reached after six iterations, with an ARI history $\{0.64,\,0.77,\,0.93,\,0.88,\,1.00\}$. The experimental set contained $N_{pts}=66$ points spanning $A_f/R \in [0.02, 0.58]$. Only five distinct regimes were identified, leading to a reduced prototype set ($n_P = 5$). The resulting clusters primarily capture stable free-surface dynamics ($\mathcal{C}_0$) and the fundamental longitudinal mode $(m=1,n=0)$ excited near twice its natural frequency ($\omega_{1,0} \approx \SI{7.12}{\radian\per\s}$).

The damping ratio for this fill level was estimated as $\zeta_{1,0} \approx 0.015 \pm 0.002$, representing an increase of roughly \SI{275}{\%} relative to $H_l/D = 0.50$. This confirms that damping increases for $H_l/D \ge 0.5$, influenced by the dome cavities and curved sidewalls (Table~\ref{tab:experimental_matrix}). This trend aligns with previous observations for spheroidal tanks, where damping exhibits a non-monotonic dependence on fill level and reaches a minimum near the point of maximum free-surface area~\citep{damping_estimation_methods_2, casiano2016damping, arndt_damping}. In contrast, canonical upright geometries (cylindrical or rectangular) typically show a leveling-off of the damping ratio beyond a critical fill height, allowing for fixed, dimensionless correlations based on the Galileo number~\citep{damping_1969_abramson, arndt_damping}. The present findings therefore deviate from these classical trends, potentially indicating additional dissipation mechanisms induced by the complex wall curvature of horizontal tanks.

Clusters $\mathcal{C}_1$ and $\mathcal{C}_2$ capture two amplitude levels of the fundamental longitudinal mode. Points near the critical forcing fall in $\mathcal{C}_1$, displaying finite-amplitude oscillations with moderate wave heights. At high fill levels, the reduced ullage volume confines the free-surface motion and enhances interfacial exchange, leading to a rapid rise in mixing between $0.1 < A_f/R < 0.2$, after which saturation occurs, as indicated by the mixing metric $I_i$ (marker size). Cluster $\mathcal{C}_2$ corresponds to strong wave-breaking at the domes with jet launches. Although these two regimes could be merged, they are kept distinct to capture the onset of breaking, which the mixing metric $I_i$ alone does not fully reveal. This separation is important in non-isothermal applications where breaking drastically affects thermodynamic behavior, causing rapid pressure fluctuations or ullage collapse~\citep{marques_experimental_2023}, and can enhance local evaporation when liquid impacts superheated walls~\citep{EUCASS25_DeMaria}.

A noteworthy transitional point appears at $A_f/R = 0.09$ and $\omega_f/(2\omega_{1,0}) = 1.00$, where the enlarged marker indicates elevated activity relative to surrounding $\mathcal{C}_0$ points. Its location near the cluster boundary reflects its borderline character. Conversely, the smallest-amplitude case in $\mathcal{C}_1$ occurs at $A_f/R \approx 0.13$, slightly above the threshold for $H_l/D = 0.50$, suggesting that finer sampling would help resolve the transition more sharply.

Higher-order responses appear in clusters $\mathcal{C}_3$ and $\mathcal{C}_4$ (right panel of Fig.~\ref{fig:final_regime_map_67}). Cluster $\mathcal{C}_3$ features a triple pulsating jet with five antinodes at $x/L = \pm 0.5$, $\pm 0.25$, and $0$, corresponding to an $m=4$-type pattern. This occurs at the map extremity under the same forcing conditions that activated the $m=3$ mode at $H_l/D = 0.50$ (Fig.~\ref{fig:final_regime_map_50}). The emergence of this $m=4$ regime is consistent with the leveling of the corresponding natural frequency at high fill ratios (Sec.~\ref{app:natfreq_theory}, Fig.~\ref{fig:natfreq_vs_fill}). Sparse sampling at high $A_f/R$ for $\omega_f/(2\omega_{1,0}) \ge 1.15$, owing to acceleration limits of the shaking table, prevented us from fully capturing the transition between the $m=4$ response in $\mathcal{C}_3$ and the longitudinal $(m=1)$ responses in $\mathcal{C}_1$ and $\mathcal{C}_4$, where time-modulated dynamics are expected.

Within $1.03 \le \omega_f/(2\omega_{1,0}) \le 1.13$, cluster $\mathcal{C}_4$ represents a transitional band between stable dynamics and the pure $(m=1)$ response. This region, delineated by dashed uncertainty contours, shows transverse motion with residual $(m=3)$ longitudinal content in the dominant spatial mode $\boldsymbol{\phi}_1$. To clarify the dynamics, Fig.~\ref{fig:67_psi_phi} presents the temporal coefficients and spectra of the first two spatial modes (the third mode being dominated by noise). The medoid exhibits similarities to prototype $p_6$ (Fig.~\ref{fig:p6_mix_high_prototype}), but without temporal modulation; instead, the system settles into a limit-cycle oscillation.

Finally, the weighting strategy in the physics-informed classifier (Subsec.~\ref{sec:SVM}) proves effective: the resulting boundaries align closely with the inviscid Mathieu prediction, as points in clusters $\mathcal{C}_1$, $\mathcal{C}_2$, and $\mathcal{C}_4$ fall near the analytical instability envelope. This agreement is remarkable given the geometric complexity of the tank, although the Mathieu model with linear damping still slightly underpredicts the excitation level required for instability.


Figure~\ref{fig:final_regime_map_40} shows the regime map for $H_l/D = 0.40 \pm 0.04$, initialized using three prototypes ($n_P=3$, $p_j = p_3$). Convergence was reached after five iterations, with an ARI history $\{0.67,\,0.65,\,0.96,\,1.00\}$. Compared to the higher-fill cases, the range of observed dynamics is markedly reduced, with stable free-surface behavior ($\mathcal{C}_0$) dominating most of the parameter space. The narrowing of the instability region is evident, particularly when contrasted with the map at $H_l/D = 0.67$. This contraction results from the overall reduction in natural frequencies, experimentally determined as $\bar{\omega}_{1,0} \approx \SI{5.32}{\radian\per\second}$. The associated instability regions for higher modes also shrink and separate more distinctly in the forcing space, leading to weaker modal interactions. Consequently, higher-order responses --- previously observed for $\omega_f/(2\omega_{1,0}) > 1.15$ at $H_l/D = 0.50$ and for $\omega_f/(2\omega_{1,0}) > 1.10$ at $H_l/D = 0.67$ --- do not arise within the explored range for this lower fill level.
 
The critical excitation amplitude appears near $A_f/R \geq 0.14$ ($A_f \geq \SI{9.4}{\mm}$), indicating that reduced liquid depth increases bottom-wall viscous dissipation and damping, thereby delaying the onset of parametric instability. This observation reinforces the need for a focused investigation into viscous dissipation mechanisms in horizontal geometries, as current literature offers limited insight into their influence on vertical sloshing dynamics.

Cluster $\mathcal{C}_1$ captures the longitudinal response and occupies the Mathieu-unstable region. At this fill ratio, the camera provides a top-view observation of the interface, introducing non-negligible three-dimensional effects that were minimal at $H_l/D = 0.50$ and $H_l/D = 0.67$.  Cluster $\mathcal{C}_2$, appearing at larger forcing amplitudes ($A_f/R \geq 0.38$) and $1.00 \leq \omega_f/(2\omega_{1,0}) \leq 1.07$, displays pronounced sloshing absent from the other classes. Its dynamics resemble prototype $p_5$ (Fig.~\ref{fig:p5_mix_low_prototype}), characterized by an $m=2$ structure with a secondary $m=1$ component, producing a mixed-mode longitudinal response. The corresponding medoid spatial modes, temporal coefficients, and spectra are shown in Fig.~\ref{fig:40_psi_phi} for further clarification.

Overall, the regimes identified in Figs.~\ref{fig:final_regime_map_50}–\ref{fig:final_regime_map_40} follow the classical behavior of parametrically excited (Faraday) waves. As shown by~\citet{benjamin_ursell_1954} and~\citet{brand_parametrically_1965}, finite onset amplitudes arise from dissipation, while the ideal Mathieu model predicts infinitesimal excitation~\citep{abramson_1981}. Consistent with these studies, the present maps exhibit the characteristic rounding of the instability tongues due to viscous and geometric effects~\citep{ciliberto_chaotic_1985, ciliberto_phenomenological_1985}. 

However, across all three fill levels, the bottom portion of the region predicted unstable by the viscous Mathieu solution remains experimentally stable. Several factors may contribute to this discrepancy. One possibility is the presence of additional dissipation mechanisms under parametric forcing --- such as amplitude-dependent viscous losses or weak nonlinear damping (with modal interaction)  or contact-line hysteresis --- not captured by the linear free-decay estimate. Yet, this explanation appears unlikely here, since the waves in this region have small amplitudes and therefore one might expected limited nonlinear content.

Another plausible explanation is that, within this band, the growth rate of the instability is extremely small, such that wave amplification does not become observable within the duration of the experiments. Nevertheless no appreciable growth trend was detected in the leading modal coefficients, indicating that --- if an instability exists --- its amplification is either too slow to emerge or effectively absent under the tested conditions. The discrepancy therefore likely reflects the combined effect of very slow (or null) parametric amplification and the simplified nature of the viscous Mathieu model, which does not capture subtle geometric or weakly nonlinear effects that may further stabilize the free surface near the stability threshold.


\section{Conclusions}\label{sec5-conclusions}

This study experimentally investigated the dynamics of vertically driven sloshing in a horizontal cylindrical tank under gravity-dominated conditions. The primary objective was to construct a dimensionless sloshing regime map and identify the boundaries between distinct flow regimes near twice the first longitudinal natural frequency, $\omega_{1,0}$ --- the principal zone of parametric instability. The results were expressed in nondimensional form using the relevant scaling laws.

Experiments were conducted at the von Kármán Institute’s SHAKESPEARE facility using a horizontal cylindrical tank with spherical end caps and aspect ratio $L/R = 5.0$. Demineralized water was used as the working fluid, and liquid fill ratios in the range $H_l/D \in [0.40,\,0.67]$ were explored over $A_f/R \in [0.02,\,0.62]$, $A_f\omega_f^2/g \in \pm[0.02,\,0.94]$, and $\omega_f/(2\omega_{1,0}) \in [0.87,\,1.20]$. This parameter sweep revealed a rich spectrum of interfacial behaviors near the primary resonance.

A data-driven methodology was introduced to construct semi-supervised regime maps by integrating unsupervised clustering and nonlinear classification within a reduced-order representation of the interface dynamics. The approach relies on a multiscale Proper Orthogonal Decomposition (mPOD) of stabilized high-speed image sequences to extract frequency-resolved modal features, which are then grouped into representative prototypes and classified using a physics-informed Support Vector Machine (SVM). This framework provides a physically interpretable mapping of sloshing behavior linked directly to the Mathieu stability analysis.

Across the tested conditions, the system exhibited transitions from stable waves to pure and mixed longitudinal modes, with oscillation amplitudes ranging from weakly nonlinear to pronounced wave breaking. Experiments confirmed that a finite excitation threshold is required to reach these regimes, consistent with the viscous Mathieu formulation. However, the experimentally observed thresholds were significantly higher than those predicted by the viscous model. This underprediction likely arises from effects not captured by the linear damping inherent to the Mathieu equation, such as extremely slow subharmonic growth rates, weak nonlinear dissipation, and modal interactions. These mechanisms effectively raise the practical stability threshold, delaying --- or suppressing --- the onset of parametric amplification within the experimental time window.

At higher fill levels ($H_l/D = 0.50$ and $0.67$), the dynamics became increasingly intricate due to competition between longitudinal modes. Mixed-mode responses emerged from the interaction between subharmonic and harmonic instability tongues associated with different sloshing modes, producing transitional regimes with amplitude modulation and intermittent switching. These behaviors are consistent with nonlinear coupling, in which energy is exchanged among near-resonant modes. The results indicate that, beyond the principal subharmonic resonance, higher-order mode interaction is a key factor governing sloshing behavior at elevated fill ratios. Future work should therefore extend the analysis to additional natural frequencies and secondary resonance bands to develop a comprehensive regime map spanning the full modal hierarchy.

Overall, this study demonstrates a novel methodology for mapping sloshing regimes using a semi-supervised image-processing pipeline that leverages modern machine-learning tools to enable scalable classification. The resulting maps highlight the potential risks associated with sloshing-induced mixing in horizontal tanks --- a geometry that has received limited attention to date. The observed mixing intensities suggest that vertical sloshing could strongly influence the tank's thermodynamic conditions, with implications for the design and operation of cryogenic storage systems.

\section{Data availability} \label{sec:data_share} 

The data supporting this study's findings will be accessible to the public upon request to the corresponding author.

\begin{acknowledgments} \label{sec:acknow} 

This work has received financial support from various sources. The experimental setup, the experimental campaign, and G. Caridi were supported by the HASTA project (EU Horizon Europe, Grant Agreement No. 101138003). Tommaso De Maria acknowledges support from the von Kármán Institute for Fluid Dynamics (VKI) through a fellowship awarded in the framework of his Research Master program. The post-processing analysis was carried out through a joint effort between Francisco Monteiro, supported by a VLAIO grant from the Flemish Government (reference HBC.2023.0897), and M. A. Mendez, supported by the European Research Council under the European Union’s Horizon Europe research and innovation programme (ERC Starting Grant RE-TWIST, Grant Agreement No. 101165479).

\end{acknowledgments}


\appendix   

\section{Analytical Solutions for Fundamental Sloshing Modes in Horizontal Cylindrical Tanks}\label{app:natfreq_theory}

A first-order approximation of the natural frequencies for sloshing modes $(m, n)$ was obtained using the methodology of \citet{hasheminejad_analytical_2017}. This approach models the tank as a perfectly rigid, finite-length, horizontal cylinder with rigid end plates, partially filled to an arbitrary depth. The analysis assumes the fluid is inviscid, incompressible, and irrotational, so the velocity potential satisfies the Laplace equation. The dynamic and kinematic boundary conditions at the free surface are linearized. Separation of variables and application of Graf’s translational addition theorem reduce the problem to an eigenvalue system for determining the natural frequencies.

\begin{figure}[b!]
\centering
\includegraphics[width=0.50\columnwidth]{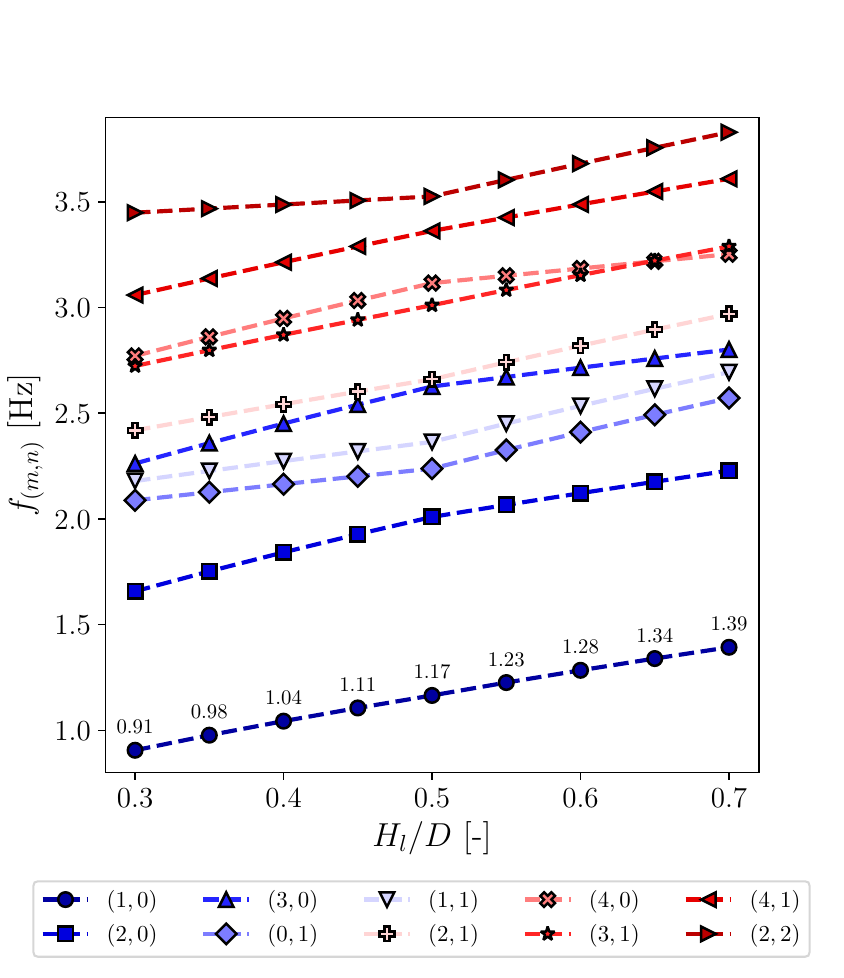}
\caption{\citet{hasheminejad_analytical_2017} derivation for the sloshing natural frequencies \(f_{(m,n)}\) [Hz] as a function of the fill level \(H_l/D\) for a horizontal cylindrical tank with flat ends with $R = \SI{67.25}{\mm}$ and aspect ratio $L/R = 5.0$. The curves are obtained by interpolation of the tabulated eigenvalues \(\Omega = 2 \pi f_{m,n}^2 R/g\) in the fill coordinate \(\bar{e} = 2H_l/D - 1\) and the aspect ratio \(L/R\). Markers denote different modal families, while annotations are reported only for the fundamental mode \((1,0)\).}
\label{fig:natfreq_vs_fill}
\end{figure}
Fig. \ref{fig:natfreq_vs_fill} presents the lowest set of modal frequencies for our tank with $R = \SI{67.25}{\mm}$ and aspect ratio $L/R = 5.0$. These frequencies are determined by interpolating tabulated eigenvalues ($\Omega$) between $L/R = 3.0$ and $L/R = 6.0$ at fill level ratios $H_l/D = 0.25$, $H_l/D = 0.50$, and $H_l/D = 0.75$. Near twice the fundamental frequency $f_{1,0}$, higher-order branches converge. This convergence results in significant overlap across various fill ratios, increasing the likelihood of modal coupling.


\bibliography{Monteiro_et_al_2025}



\end{document}